# Direct Frequency Comb Measurement of OD + CO → DOCO Kinetics

**Authors:** B. J. Bjork[1], T. Q. Bui[1], O. H. Heckl[1], P. B. Changala[1], B. Spaun[1], P. Heu[2], D. Follman[2], C. Deutsch[3], G. D. Cole[2,3], M. Aspelmeyer[4], M. Okumura[5], J. Ye[1]

**Affiliations:**

[1]JILA, National Institute of Standards and Technology and University of Colorado, Department of Physics, University of Colorado, Boulder, CO 80309, USA

[2]Crystalline Mirror Solutions LLC, 114 E Haley St., Suite G, Santa Barbara, CA 93101, USA

[3]Crystalline Mirror Solutions GmbH, Parkring 10, 1010 Vienna, Austria

[4]Vienna Center for Quantum Science and Technology (VCQ), Faculty of Physics, University of Vienna, 1090 Vienna, Austria

[5]Arthur Amos Noyes Laboratory of Chemical Physics, Division of Chemistry and Chemical Engineering, California Institute of Technology, 1200 East California Boulevard, Pasadena, CA 91125, USA

*Correspondence to: bryce.bjork@colorado.edu, Ye@jila.colorado.edu

**Abstract**: The kinetics of the OH + CO reaction, fundamental to both atmospheric and combustion chemistry, are complex due to the formation of the HOCO intermediate. Despite extensive studies on this reaction, HOCO has not been observed at thermal reaction conditions. Exploiting the sensitive, broadband, and high-resolution capabilities of time-resolved cavity-enhanced direct frequency comb spectroscopy, we observe OD + CO reaction kinetics with the detection of stabilized *trans*-DOCO, the deuterated analogue of *trans*-HOCO, and its yield. By simultaneously measuring the time-dependent concentrations of both *trans*-DOCO and OD species, we observe unambiguous low-pressure termolecular dependence on the reaction rate coefficients for both $N_2$ and CO bath gases. These results confirm the HOCO formation mechanism and quantify its yield.

**One Sentence Summary:** We detect *trans*-DOCO and OD in the reaction of OD + CO using cavity-enhanced direct frequency comb spectroscopy and determine the kinetics and *trans*-DOCO branching yield in the low pressure regime.

**Main Text:**

The apparent simplicity of gas phase bimolecular reaction kinetics of free radicals often belies the complexity of the underlying dynamics. Reactions occur on multidimensional potential energy surfaces that can possess multiple pre-reactive complexes, bound intermediate complexes and multiple transition states. As a result, effective bimolecular rate coefficients often exhibit complex temperature and pressure dependence. The importance of free radical reactions in processes such as combustion and air pollution chemistry has motivated efforts to determine these rate constants both experimentally and theoretically. Quantitative *ab initio* modeling of kinetics remains a major contemporary challenge (*1*), requiring accurate quantum chemical calculations of energies, frequencies and anharmonicities, master equation modeling, energy transfer dynamics, and, when necessary, calculation of tunneling and non-statistical behavior. Experimentally, detection of the transient intermediates, which is the key to unraveling the dynamics, is often quite challenging.

The reaction,

$$OH + CO \rightarrow H + CO_2\,, \qquad \Delta H_0^{o} = -103.29 \text{ kJ/mol}, \qquad (1)$$

has been extensively studied over the last four decades because of its central role in atmospheric and combustion chemistry (*2*); it has come to serve as a benchmark for state-of-the-art studies of chemical kinetics of complex bimolecular reactions (*3*, *4*). In Earth's atmosphere, the hydroxyl radical OH is critical as the primary daytime oxidant (*5*, *6*). CO, a byproduct of fossil fuel burning, acts through reaction 1 as an important global sink for OH radicals and is the dominant OH loss process in the free troposphere. In fossil fuel combustion, OH + CO is the final step that oxidizes CO to $CO_2$ and is responsible for a large amount of heat released.

The rate of reaction 1 is pressure dependent and exhibits an anomalous temperature dependence, which led Smith and Zellner (*7*) to propose that the reaction proceeds through a highly energized, strongly bound intermediate, HOCO, the hydrocarboxyl radical.

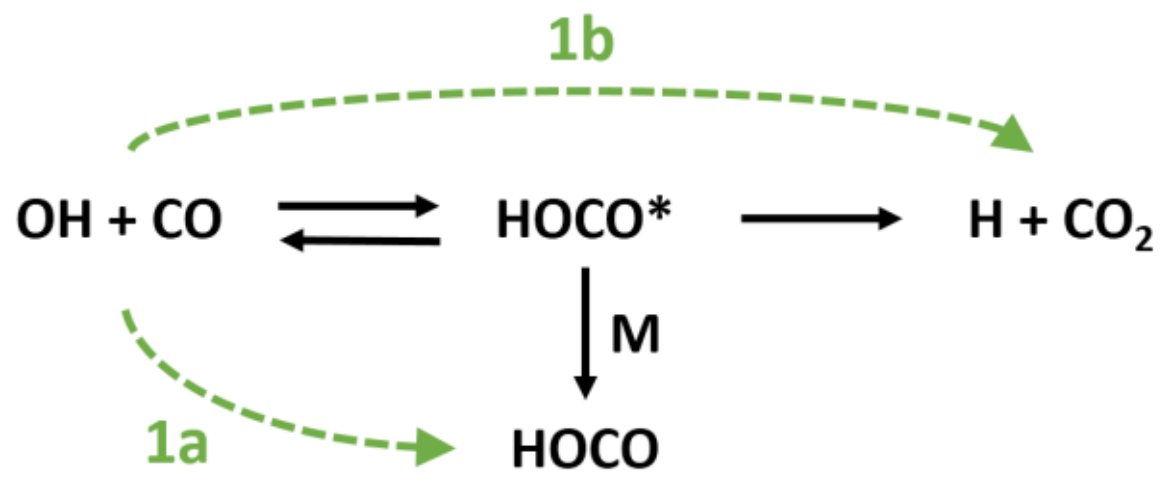

(2)

Formation of H + $CO_2$ products is an example of a chemically activated reaction. The course of the reaction is governed by the dynamics on the potential energy surface, shown schematically in Fig. 1A. The OH and CO pass through a pre-reactive weakly bound OH-CO complex to form a highly energized HOCO* in one of two isomers, *trans*-HOCO or the less stable *cis*-HOCO. In the low pressure limit at room temperature, HOCO* primarily back-reacts to OH + CO, but there is a small probability to overcome the low barrier (8.16 kJ/mol) to react to form H + $CO_2$. In the presence of buffer gas, energy transfer by collisions with third-bodies M (termolecular process) can deactivate or further activate the HOCO*. Deactivation can lead to the formation of stable, thermalized HOCO products (reaction 1a in Scheme 2) and reduces the formation of H + $CO_2$ (reaction 1b in Scheme 2). As one approaches the high pressure limit, HOCO formation becomes the dominant channel and H + $CO_2$ product formation decreases. The overall reaction rate is characterized by an effective bimolecular rate constant $k_1([M],T) = k_{1a}([M],T) + k_{1b}([M],T)$ (*8-12*).

There have been numerous experimental studies of the temperature and pressure dependence of the overall rate coefficient $k_1([M],T)$; these all measure OH loss in the presence of CO (*9, 11-17*). In principle, one can use master equation calculations with accurate potential energy surfaces within a statistical rate theory to compute $k_1([M],T)$, but *a priori* kinetics are rarely possible because the energy transfer dynamics are generally not known. A number of studies have thus fit the theoretical models to the observed overall rate constants, using a small number of parameters to describe collisional energy relaxation/activation (*9, 11, 15, 16, 18, 19*). While these previous studies have had success in describing $k_1([M],T)$, they do not capture the dynamics that would be revealed from the pressure-dependent branching between stabilization of HOCO and barrier crossing to form H + $CO_2$ products. Detection of the stabilized HOCO intermediate and measurement of its pressure-dependent yield would confirm the reaction mechanism and quantitatively test theoretical models. The spectroscopy of HOCO is well established, and recently HOCO has been observed from the OH + CO reaction generated in a discharge (*20-22*); however one cannot derive rate constants from non-thermal conditions.

To directly and simultaneously measure the time-dependent concentrations of reactive radical intermediates such as HOCO and OH, we use a recently developed technique of time-resolved direct frequency comb spectroscopy (TRFCS) (*23*). The massively parallel nature of frequency-comb spectroscopy allows for time-resolved, simultaneous detection of a number of key species, including intermediates and primary products, with high spectral and temporal resolution. The light source is a mid-IR ($\lambda \approx 3$-$5$ μm) frequency comb, generated from an optical parametric oscillator (OPO) synchronously pumped with a high repetition rate ($f_{rep}$ = 136 MHz) mode-locked femtosecond fiber laser (*24*). The OPO spectrum is composed of spectrally narrow "comb teeth" that are evenly spaced by $f_{rep}$, and shifted by an offset frequency, $f_0$. By matching and locking the free spectral range (FSR) of the enhancement cavity to $2 \times f_{rep}$, the full comb spectrum remains resonant with the cavity during the data acquisition. The broadband transmitted light (~65 $cm^{-1}$ bandwidth, ~7100 comb teeth) is spatially dispersed in 2D by a VIPA etalon and a grating combination, which is then imaged on an InSb camera (Fig. 1C). Absorption spectra are constructed from these images as a function of time (with a resolution of ≥10 μs determined by the camera integration time), which are compared with known molecular line intensities to obtain absolute concentrations. The absorption detection sensitivity is greatly enhanced with our high finesse (F≈4100) optical cavity that employs mid-IR mirrors with low-loss crystalline coatings. These mirrors, with a center wavelength of 3.72 μm and a spectral bandwidth of about 100 nm, have significantly lower optical losses and hence yield enhanced cavity contrast compared with traditional amorphous coatings (*25*), enabling an improved sensitivity by a factor of 10 for the direct detection of *trans*-DOCO.

In this experiment, we have studied the deuterium analogue of reaction 1, OD + CO → D + $CO_2$, exploiting the sensitivity and resolution of TRFCS to detect the reactant OD (in both v=0 and v=1 states) and the product *trans*-DOCO by absorption spectroscopy in a pulsed-laser-photolysis flow cell experiment. We sought to measure the pressure-dependent effective bimolecular rate coefficients and the yield of *trans*-DOCO at 27-75 Torr total pressure. Such measurements would be especially sensitive to the competition between termolecular DOCO stabilization and reaction to form D+$CO_2$. Detection of the deuterated species allowed us to avoid atmospheric water interference in our spectra. We further anticipated that the yield of stable DOCO would be higher, since deuteration significantly reduces the rate of tunneling to form D + $CO_2$ products while increasing the lifetime of DOCO* due to the higher density of states.

The OD + CO reaction was initiated in a slow-flow cell by photolyzing $O_3$ in a mixture of $D_2$, CO, and $N_2$ gases with 266 nm (32 mJ) pulses from a frequency-quadrupled Nd:YAG laser, expanded to a profile of 44 mm × 7 mm and entering the cell perpendicular to the cavity axis. $[O_3]_0$ was fixed at a starting concentration of $1\times10^{15}$ molecules $cm^{-3}$ and verified by direct UV absorption spectroscopy. The initial concentrations of CO, $N_2$, and $D_2$ were varied over the range 1 – 47 Torr while the $O_3$ concentration was restricted to 3 – 300 mTorr to minimize secondary reactions. A complete description and tabulation of the experimental conditions is included in the supplementary information (SI, **§1**).

Each photolysis pulse dissociated 15% of the ozone (SI, **§1**) to form $O_2$ + $O(^1D)$ at nearly unity quantum yield (*26*). The resulting $O(^1D)$ either reacts with $D_2$ to form OD + D or is quenched by background gases to $O(^3P)$ within 1 μs. $O(^1D)$ + $D_2$ is known to be highly exothermic and produces vibrationally excited OD(v = 0 - 4) with an inverted population peaking at v = 2 and v = 3 (*27*). Vibrationally excited OD was rapidly quenched or formed D atoms by collisions with CO (*28, 29*). Formation of vibrational Feshbach resonances of DOCO* from collisions of OD(v>0) with CO may be possible, but the lifetimes are on the order of picoseconds, as previously observed for the HOCO* case (*30-33*). Therefore, only vibrationally/rotationally thermalized OD(v=0) are expected to form DOCO by the mechanism described in Scheme 2. OD and DOCO reach a steady-state after 100 μs through cycling reactions depicted in Fig. 1B: D atoms produced from OD + CO $\rightarrow$ D + $CO_2$ reacted with $O_3$ to regenerate the depleted OD.

Absorption spectra covering a ~65 $cm^{-1}$ bandwidth were recorded at a sequence of delays from the $t$ = 0 photolysis pulse, using a camera integration time of either 10 or 50 μs depending on our sensitivity to *trans*-DOCO signals. The broad bandwidth of the comb covers 6 OD, ~200 $D_2O$, and ~150 *trans*-DOCO transitions. These spectra were normalized to a spectrum acquired directly preceding the photolysis pulse and were fitted to determine time-dependent concentrations. With this approach, we captured the time-dependent kinetics of *trans*-DOCO, OD, and $D_2O$ from OD + CO within a spectral window of 2660-2710 $cm^{-1}$. Representative snapshots at three different delay times are shown in Fig. 2A. The OD and *trans*-DOCO data were compared to simulated spectra, generated with PGopher (*34*) using measured molecular constants (*35-37*) and known or computed intensities. The simulated spectra are fitted to these experimental data at each time delay to map out the full time trace of the three observed species, as shown in Fig. 2B-C, with error bars derived directly from the fit residual. See (SI, **§2**) for details of data analysis.

We determined the effective bimolecular rate coefficient for the *trans*-DOCO channel, $k_{1a}$([M],T) from simultaneous measurements of both time-dependent *trans*-DOCO and OD. In the low pressure regime studied here, the DOCO formation rate obeys a termolecular rate law, while the effective bimolecular coefficient for the D + $CO_2$ channel remains close to the zero pressure value, $k_{1b}$([M]=0). We measured the dependence of the effective bimolecular rate constant on the concentrations of all of the major species ($N_2$, CO, $D_2$, and $O_3$) present in the experiment.

We analyzed the early-time ($t$ <200 μs) rise of *trans*-DOCO in order to decouple the measurement of $k_{1a}$ from secondary loss channels at longer times. The expected time dependence of the DOCO concentration is given by

$$\frac{\mathrm{d[DOCO]}}{dt} = k_{\mathrm{1a}}\left[\mathrm{CO}\right]\left[\mathrm{OD}\right](t) - k_{\mathrm{loss}}\left[\mathrm{X}\right]\left[\mathrm{DOCO}\right](t). \tag{3}$$

$k_{\mathrm{loss}}$ describes a general DOCO decay through a reaction with species X and [OD]($t$) refers to the time-dependent concentration of OD in the ground vibrational state. The solution to Eq. 3 is a convolution of the DOCO loss term with [OD](t), given by the integral in Eq. 4. [CO] is in large excess and remains constant throughout the reaction.

$$[\mathrm{DOCO}](t) = k_{\mathrm{1a}}[\mathrm{CO}]\int_0^t e^{-(k_{\mathrm{loss}}[\mathrm{X}])(t-u)}[\mathrm{OD}](u)du$$

(4)

The effective bimolecular rate coefficient $k_{1a}$ can be reduced into two terms dependent on $N_2$ and CO concentrations,

$$k_{1a} = k_{1a}^{(CO)}\left[\mathrm{CO}\right] + k_{1a}^{(N2)}\left[\mathrm{N_2}\right] \ , \tag{5}$$

where $k_{1a}^{(CO)}$ and $k_{1a}^{(N2)}$ are the termolecular rate coefficient dependence on CO and $N_2$, respectively.

By simultaneously fitting [DOCO]($t$) and [OD]($t$) as a function of [CO] and [$N_2$], we uniquely determined all of the $k_{1a}$ termolecular coefficients. Figure 2B shows an early-time segment of our data at 10 μs camera integration for both [*trans*-DOCO]($t$) and [OD]($t$). To fit the nonlinear time-dependence of [OD]($t$), we use derived analytical functions comprised of the sum of boxcar-averaged exponential rise and fall functions (SI, **§3**). Equation 4 gives the functional form for fitting [*trans*-DOCO]($t$), which includes the integrated [OD]($t$) over the fitted time window of -25 to 160 μs. The fitted parameters are $k_{1a}$ and a *trans*-DOCO loss rate, $r_{loss,exp}$ (≡ $k_{loss}$[X]).

For our first set of data, we varied the CO concentration. At each set of conditions, we acquired data at both 10 and 50 μs camera integration time. By plotting $k_{1a}$ versus [CO] for both 10 μs and 50 μs, we did not observe any systematic dependence on camera integration time. Moreover, we observed a clear linear dependence, indicating a strong termolecular dependence of $k_{1a}$ on CO, or $k_{1a}^{(CO)}$ (Fig. 3A). The offset in the linear fit comes from the $N_2$ termolecular dependence of $k_{1a}$, or $k_{1a}^{(N2)}$. We then varied $N_2$ concentration and observed a similar linear dependence of $k_{1a}$ from Eq. 5. A 50 μs camera integration time was used for this second data set due to lower *trans*-DOCO signals at higher $N_2$ concentrations. The results are shown in Fig. 3B. Since both the offset terms from the linear fit to the CO data and the linear fit to the $N_2$ plot correspond to $k_{1a}^{(N2)}$, we performed a multidimensional linear regression to Eq. 3 to determine $k_{1a}^{(CO)}$, $k_{1a}^{(N2)}$, and $r_{loss}$ simultaneously. Since, $r_{loss,exp}$ describes *trans*-DOCO loss, it is expected to be invariant to [CO] and [$N_2$]. Therefore, $r_{loss,exp}$ serves as a shared, fitted constant in the global fit across the CO and $N_2$ data sets. From the fits shown in red in Fig. 3A-B, we obtained $k_{1a}^{(N2)} = (1.1 \pm 0.4) \times 10^{-32}$ cm$^6$ molecules$^{-2}$ s$^{-1}$, $k_{1a}^{(CO)} = (1.7 \pm 0.7) \times 10^{-32}$ cm$^6$ molecules$^{-2}$ s$^{-1}$ and $r_{loss,exp} = (4.0 \pm 0.4) \times 10^4$ s$^{-1}$. The statistical and systematic errors in these parameters are obtained from the fitting procedure and are discussed in (SI, **§5)**.

To verify the reaction kinetics, we constructed a rate equation model of the OD + CO chemistry, which included the decay channels from secondary chemistry, in order to fit the *trans*-DOCO and OD time traces up to 1 ms (SI, **§4)**. We fit one overall scaling factor for both OD and *trans*-DOCO, which accounts for uncertainties in (1) the optical path length and (2) photolysis yield and subsequent OD* quenching reactions that establish the initial steady-state concentration of OD. We also fit an additional *trans*-DOCO loss, $r_{loss,model}$, in order to correctly capture the *trans*-DOCO concentration at t >100 μs.

The *trans*-DOCO + $O_3 \rightarrow$ OD + $CO_2$ + $O_2$ rate coefficient (*9*) ($k_{O3+DOCO} \approx 4 \times 10^{-11}$ cm$^3$ molecules$^{-1}$ s$^{-1}$) and the OD + CO termolecular rate coefficients from our experimentally measured values were fixed in the model. Representative fits for two different conditions using the same rate equation model are shown in Fig. 4A-B. We found good fits with a single, consistent set of parameters over a wide range of CO, $N_2$, and $O_3$ concentrations, giving $r_{\text{loss,model}}$ = $(4.7\pm0.7)\times10^3$ $s^{-1}$ for all conditions (Fig. S10A). The sum of loss contributions from $k_{O3+DOCO}[O_3]$ and an additional loss from $r_{\text{loss,model}}$ gives a total loss of ~$4.5 \times 10^4$ $s^{-1}$, consistent with our measured $r_{\text{loss,exp}}$. One possibility for $r_{\text{loss,model}}$ is a second product branching channel of *trans*-DOCO + $O_3$ to produce $DO_2$ + $CO_2$ + O. The slight discrepancy of the *trans*-DOCO data with the rate equation model in Fig. 4B is possibly due to the inadequately constrained loss processes at long delay times.

Sources of systematic uncertainty have been carefully evaluated. First, we considered the impact of vibrationally hot OD at early times. We constrained the population of vibrationally excited OD in our system by directly observing several hot band transitions from OD(v=1) (Fig. S5). We observed that CO is an efficient quencher of OD vibration, with a measured OD(v=1) lifetime (Fig. S6) consistent with the OD(v=1) + CO quenching rate reported by Brunning *et al.* (*17*) and Kohno *et al.* (*29*). These measurements reveal that the lifetime is well below the minimum integration time of 10 μs and the total population of [OD(v=1)] is less than 10% of [OD(v=0)] in this time window. As OD(v=1) is expected to produce stabilized *trans*-DOCO less efficiently than OD(v=0), the systematic effect caused by the vibrationally hot OD is estimated to be < 10%, which has been included in our total error budget (SI, §5).

Another systematic uncertainty arises from the finite camera integration time, which is large (50 μs) or comparable (10 μs) to the early *trans*-DOCO rise time. The recovered $k_{1a}$ values from the two integration times are consistent with each other to within 21%, which we have included as a systematic uncertainty in our measurement (Fig. S4).

A third source of systematic uncertainty comes from any factors that would cause deviations from Eq. 3; specifically, we investigated the dependence of $k_{1a}$ on $D_2$ and $O_3$ concentrations. Additional experiments were conducted in the same manner as the CO and $N_2$ experiments but varying $[O_3]$ ($1\times10^{14}$ – $4\times10^{15}$ molecules $cm^{-3}$) and $[D_2]$ ($7\times10^{16}$ – $1\times10^{18}$ molecules $cm^{-3}$). At our experimental conditions and using a 50 μs camera integration window, we observed a weak dependence of $k_{1a}$ on $[O_3]$ (Fig. S7) and no statistically significant variation with $[D_2]$ (Fig. S8). The $O_3$ dependence was measured at a CO concentration of $1.5\times10^{17}$ molecules $cm^{-3}$. From analysis of the early-time *trans*-DOCO rise as a function of $[O_3]$ and $[D_2]$, we determined that $O_3$ and $D_2$ contributes an additional 11% and 8% statistical uncertainty, respectively, to our total budget. Table S3 includes the full evaluation of all known systematic uncertainties.

We find that CO is approximately 50% more effective as a collision partner than $N_2$ in promoting the termolecular association of *trans*-DOCO. This result was missed in previous experiments, which minimized the CO concentration ($<4\times10^{16}$ molecules $cm^{-3}$) to avoid biasing their pseudo-first order kinetics measurement (*12, 38*). One might naively expect CO to be similar to $N_2$ as a third body; the significant difference observed here could be due to (1) near-resonant energy transfer between CO and the CO mode in DOCO, (2) a stronger interaction potential between CO and DOCO*, or (3) the influence of higher CO on OD(v) quenching which we have not correctly accounted for.

In the low pressure regime, our measurements of the association rate, $k_{1a}$, can be compared to the pressure dependence of $k_1$, the overall rate of OD+CO, measured in previous experiments in $N_2$. Most of the pressure dependence of $k_1$ comes from $k_{1a}$, because $k_{1b}$ is expected to change only slightly in this range. The termolecular dependence of $k_1$, derived from fitting the data of Paraskevopoulos *et al*. (*14*) and Golden *et al*. (*11*) are, $k_1^{(N2)} = 1.1\times10^{-32}$ and $8.2\times10^{-33}$ $cm^6$ $molecules^{-2}$ $s^{-1}$, respectively, in reasonably good agreement with our results for $k_{1a}$. Apparent curvature in the pressure dependence seen elsewhere suggests that $k_{1a}$ may already be in the fall-off regime. In order to estimate the *trans*-DOCO branching yield (% yield $\approx k_{1a}/(k_{1a} + k_1(p=0))$, we took the average value of $k_1$ from Paraskevopoulos *et al.*, Golden *et al*., and Westernberg *et al*. (*39*). Even at low total pressures (75 Torr of $N_2$), our results show that OD+CO produces a *trans*-DOCO yield of nearly (32±10)%.

Optical frequency comb spectroscopy provides broad-band, time-resolved absorption detection of radicals with exceptional sensitivity and high spectral resolution. These results clearly demonstrate the capabilities of time-resolved cavity-enhanced frequency comb spectroscopy to elucidate chemical mechanisms through the quantitative detection of intermediates and primary products in real time. Our quantification of the termolecular dependence reveals additional factors that impact the product branching of the OH + CO reaction, which must be included in future atmospheric and combustion model predictions. For example, sensitivity analyses by Boxe *et al*. (*40*) have shown that depending on the branching ratio, HOCO can contribute 25-70% of the total $CO_2$ concentration in the Martian atmosphere. The current experiment can be readily extended to detect other primary products ($DO_2$, $CO_2$) as well as to study the OH/HOCO system. Furthermore, dynamics and nonthermal processes such as chemical activation, energy transfer and rovibrational state-specific kinetics can be studied. With the bandwidth of optical frequency comb sources spanning an octave or more, the potential of this approach has not yet been fully realized. The technologies of frequency comb sources, detection methods and mirror coatings are developing rapidly and will allow for more expansive applications of this multiplexed technique to many other classes of important chemistry problems.

**Figure (1):** **(A)** OH + CO → H + $CO_2$ potential energy surface, with energies taken from Nguyen *et al*. (*2*). OH + CO → H + $CO_2$ proceeds through vibrationally excited HOCO*, which is either deactivated by bath gas M or reacts to form H + $CO_2$. **(B)** Reaction schematic indicating the most important reactions in our system. Time-dependent concentrations of *trans*-DOCO, OD(v=0), OD(v=1), and $D_2O$ (red) are measured by cavity-enhanced absorption spectroscopy, while the concentrations of the precursors (purple) are set by flow controllers/meters. $O_3$ is measured by UV absorption. **(C)** A mid-IR frequency comb is coupled into an enhancement cavity, consisting of two high reflectivity crystalline (HRc) mirrors, where a 266 nm laser pulse photolyzes $O_3$ to initiate the chemistry. The transmission from the cavity is spatially dispersed by a VIPA etalon and a diffraction grating and imaged on an InSb camera. Simulated cavity absorbance images are shown for OD (red), *trans*-DOCO (green), and $D_2O$ (magenta) to illustrate the camera pixel to wavelength mapping.

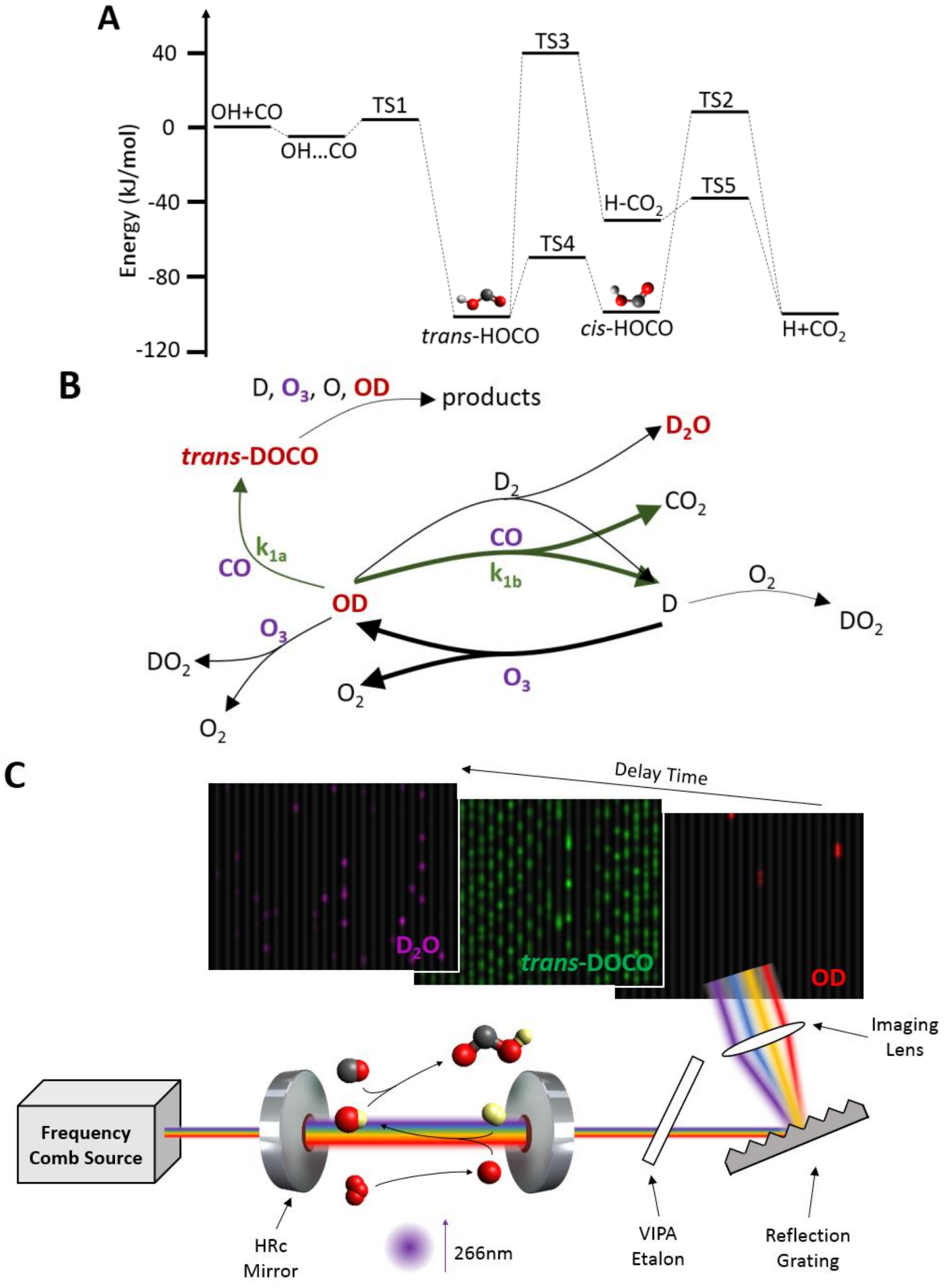

**Figure (2): (A)** Spectral fitting: Experimental spectra (black) are recorded with an integration time of 50 μs and offsets of -50 ("Before Photolysis"), 100, and 4000 μs from the photolysis pulse. These spectra are then fitted to the known line positions of OD (blue), $D_2O$ (green), and *trans*-DOCO (orange) to determine their temporal concentration profiles. The P, Q, and R-branches of *trans*-DOCO are indicated above the 100 μs experimental trace. **(B)** An analytical functional form for OD(t) is obtained by fitting the data (black circles) to a sum boxcar-averaged exponential functions (red line). **(C)** The bimolecular *trans*-DOCO rise rate is obtained by fitting the data (black circles) to Eq. 4 (red line). The data in Fig. 2B-C are obtained at a 10 μs camera integration time and precursor concentrations [CO] = $5.9\times10^{17}$, [$N_2$] = $8.9\times10^{17}$, [$D_2$] = $7.4\times10^{16}$, and [$O_3$] = $1\times10^{15}$ molecules cm$^{-3}$.

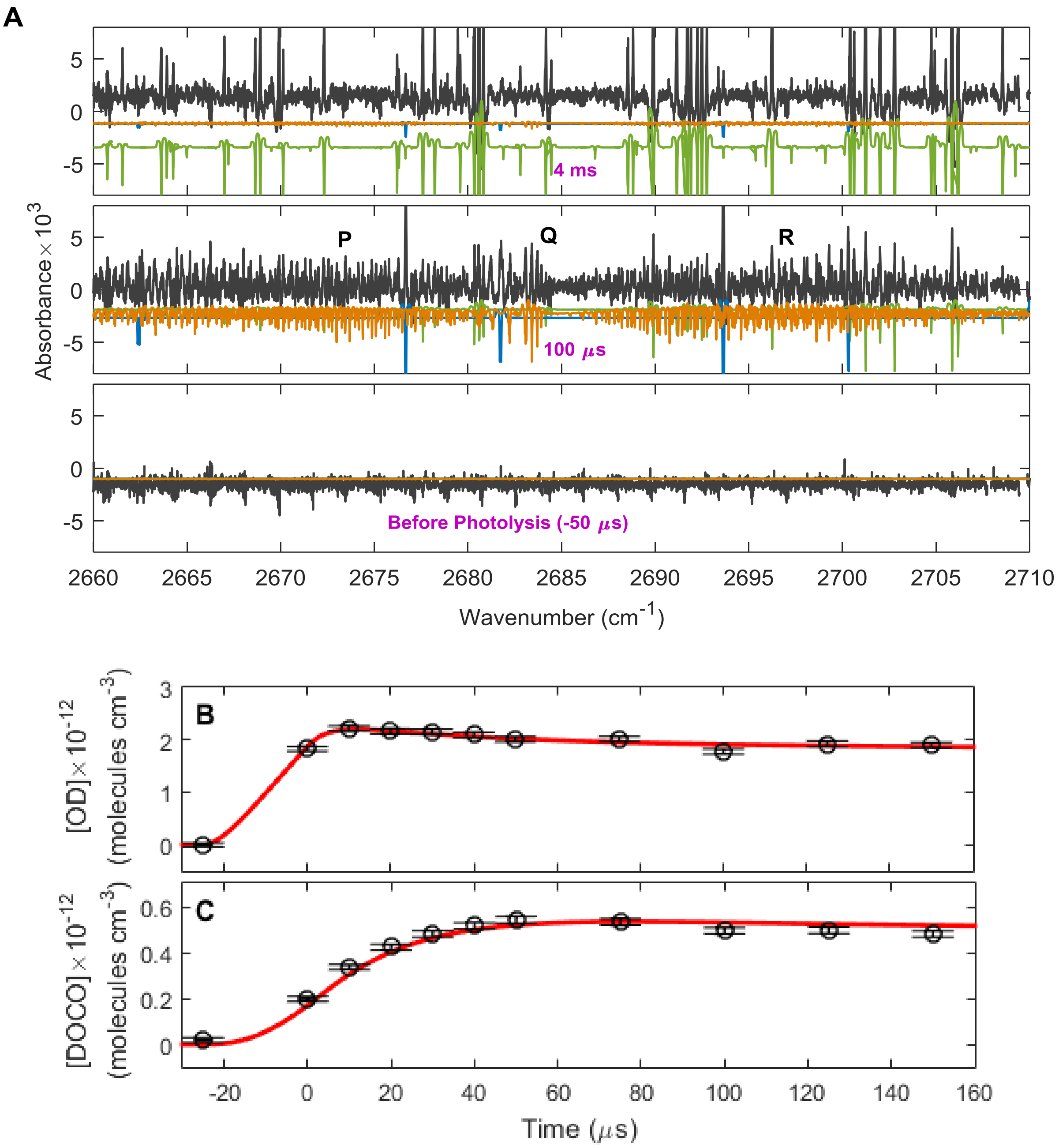

**Figure (3):** Determination of the termolecular *trans*-DOCO formation rate. The bimolecular *trans*-DOCO formation rate, $k_{1a}$, is plotted as a function of [CO] and [$N_2$] to determine the termolecular rate coefficients $k_{1a}^{(CO)}$ and $k_{1a}^{(N2)}$. Each point represents one of 26 experimental conditions tabulated in Table S1. **(A)** $k_{1a}$ is plotted as a function of [CO] while [$N_2$]=8.9×10$^{17}$ molecules cm$^{-3}$ is held constant. **(B)** $k_{1a}$ is plotted as a function of [$N_2$] while [CO] = 5.6×10$^{17}$ molecules cm$^{-3}$ is held constant. In all plots, $D_2$ and $O_3$ concentrations are fixed at [$D_2$] = 7.4×10$^{16}$ molecules cm$^{-3}$ and [$O_3$] = 1×10$^{15}$ molecules cm$^{-3}$. Red and blue data points indicate 50 and 10 μs camera integration time, respectively. The data in (**A**), (**B**) are simultaneously fit to Eq. 5. The black lines in (**A**), (**B**) are obtained from linear fits. The y offsets in the data arise from the non-zero concentrations of $N_2$ and CO in (**A**) and (**B**), respectively.

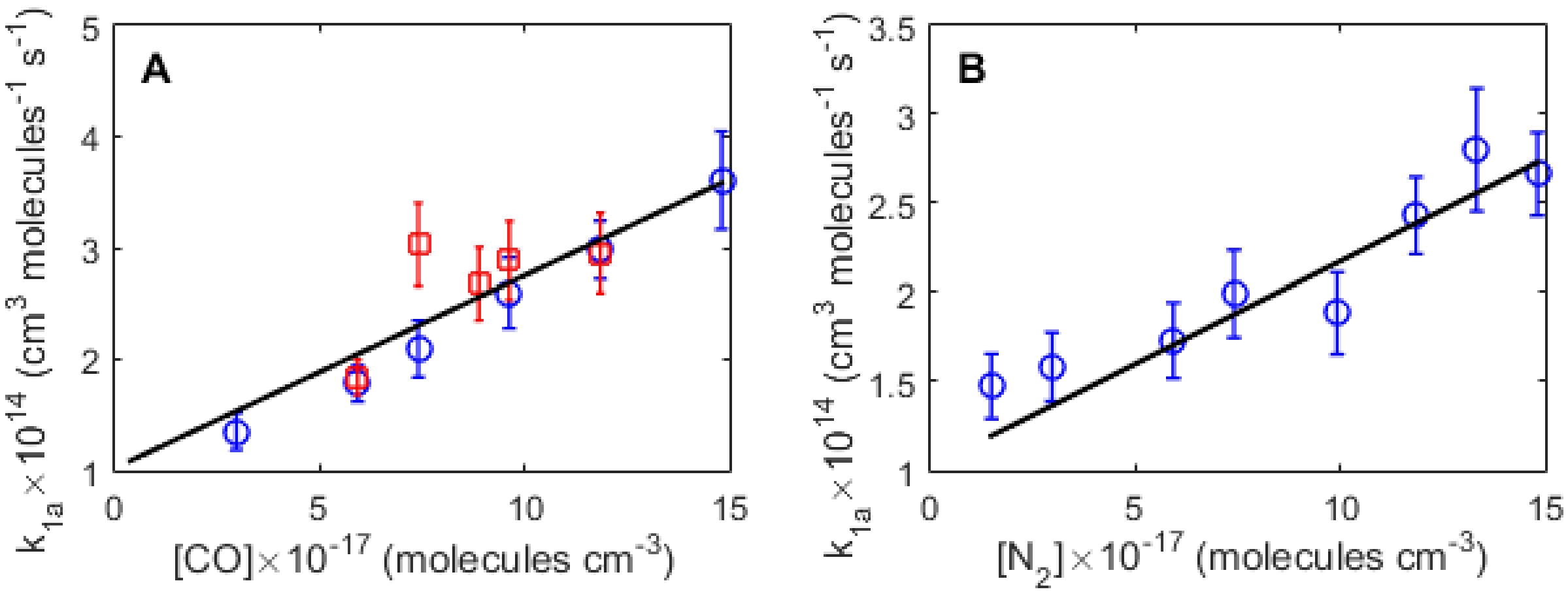

**Figure (4):** The OD (blue) and *trans*-DOCO (red) traces are fitted to the full rate equation model (solid and dashed lines for OD and *trans*-DOCO, respectively), described in the supplementary materials. Here, the integration time is 50 μs. The input $k_{1a}$ values for both CO and $N_2$ are from the early-time *trans*-DOCO rise analysis and are fixed in the fit. The floating parameters are a single scaling factor for the OD and *trans*-DOCO intensities and an extra DOCO loss channel. **(A**) [CO] = $5.9\times10^{17}$ molecules $cm^{-3}$; **(B)** [CO] = $1.2\times10^{18}$ molecules $cm^{-3}$. [$N_2$] = $8.9\times10^{17}$, [$D_2$] = $7.4\times10^{16}$, and [$O_3$] = $1\times10^{15}$ molecules $cm^{-3}$ are fixed for both data sets.

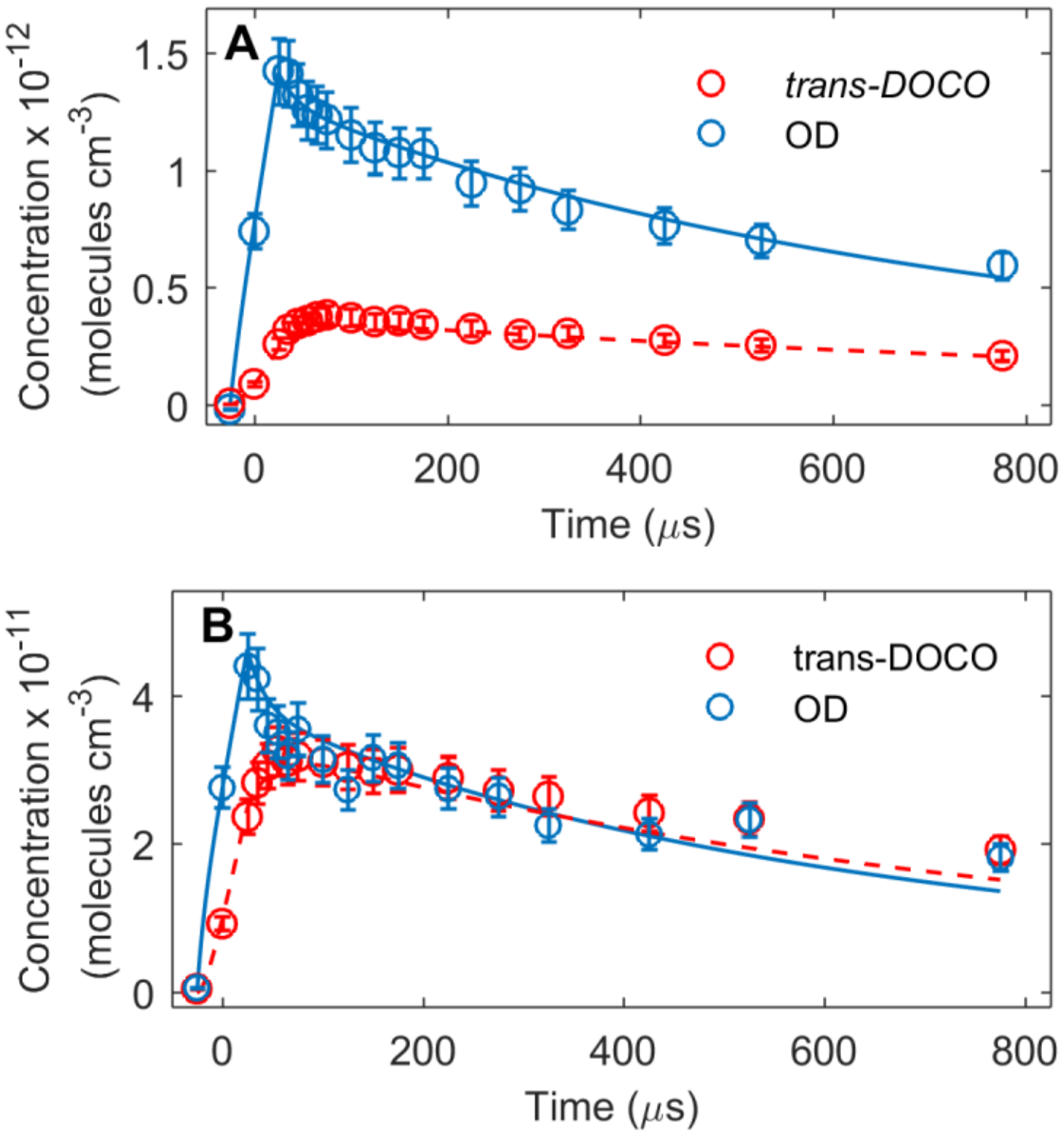

**Acknowledgments:** We thank Keeyoon Sung of the National Aeronautics and Space Administration (NASA) Jet Propulsion Laboratory (JPL) for providing a list of $D_2O$ mid-IR line positions and intensities measured by Robert A. Toth of JPL. We acknowledge financial support from AFOSR, DARPA SCOUT, NIST, NSF, and DARPA (FAA-9550-14-C-0030 and W31P4Q-16-C-0001). T. Q. Bui and B. Spaun are supported by the National Research Council Research Associate Fellowship, P. B. Changala is supported by the NSF GRFP, and O.H. Heckl is partially supported through a Humboldt Fellowship.

**Supplementary Materials**

Supplementary Materials and Methods

Figures S1-S10

Tables S1-S3

References (*S1*) – (*S31*)

Supplementary Material for

# Direct Frequency Comb Measurement of OD + CO → DOCO Kinetics

**Authors:** B. J. Bjork[1], T. Q. Bui[1], O. H. Heckl[1], P. B. Changala[1], B. Spaun[1], P. Heu[2], D. Follman[2], C. Deutsch[3], G. D. Cole[2,3], M. Aspelmeyer[4], M. Okumura[5], J. Ye[1]

**Affiliations:**

[1]JILA, National Institute of Standards and Technology and University of Colorado, Department of Physics, University of Colorado, Boulder, CO 80309, USA

[2]Crystalline Mirror Solutions LLC, 114 E Haley St., Suite G, Santa Barbara, CA 93101, USA

[3]Crystalline Mirror Solutions GmbH, Parkring 10, 1010 Vienna, Austria

[4]Vienna Center for Quantum Science and Technology (VCQ), Faculty of Physics, University of Vienna, 1090 Vienna, Austria

[5]Arthur Amos Noyes Laboratory of Chemical Physics, Division of Chemistry and Chemical Engineering, California Institute of Technology, 1200 East California Boulevard, Pasadena, CA 91125, USA

*Correspondence to: bryce.bjork@colorado.edu, Ye@jila.colorado.edu

**CONTENTS:**

**Supplementary Materials and Methods:**

§1. Experimental description and conditions

***Generation of OD:*** The ozone used to generate OD in this reaction was generated in a flow-discharge of pure $O_2$ gas. This mixture contains approximately 8% of $O_3$ in a buffer of $O_2$. This mixture was flown across a silica gel trap immersed in an isopropanol/$LN_2$ bath at -90 °C. The ozone trap was allowed to pump out for about 20 minutes after stopping the $O_3/O_2$ flow to remove residual $O_2$. The steady state concentration of $O_3$ in the reaction cell was measured using the direct absorption of the collimated 270 nm light from a UV LED (UVTOP-270). By comparing this absorption measurement to a static total pressure measurement in the cell, it is estimated that $O_3$ comprises >70% of the mixture flowing. Frequency quadrupled 266 nm light (beam size: 44 mm × 7 mm, power = 32mJ/pulse) from a Spectra Physics INDI-HG-105 Nd:YAG propagating orthogonal to the mid-IR probe beam was the photolysis beam. In this method, 15% of the $O_3$ in the cavity was photolyzed into $O(^1D)$, $O(^3P)$, and $O_2$ to start the reaction. OD was then promptly formed from the $O(^1D) + D_2 \rightarrow OD + D$ reaction.

***$O_3$ photolysis to $O(^1D)$:*** In order to accurately simulate the kinetics of the OD + CO chemical system, we measured the fraction of $O_3$ photolyzed at 266 nm. This was done by measuring the transmission of a 270 nm LED through the photolysis region of the chemical cell before and after the photolysis of $O_3$ in a buffer of $N_2$. We measured photolysis fraction of $f_{phot} = 0.15 \pm 0.02$.

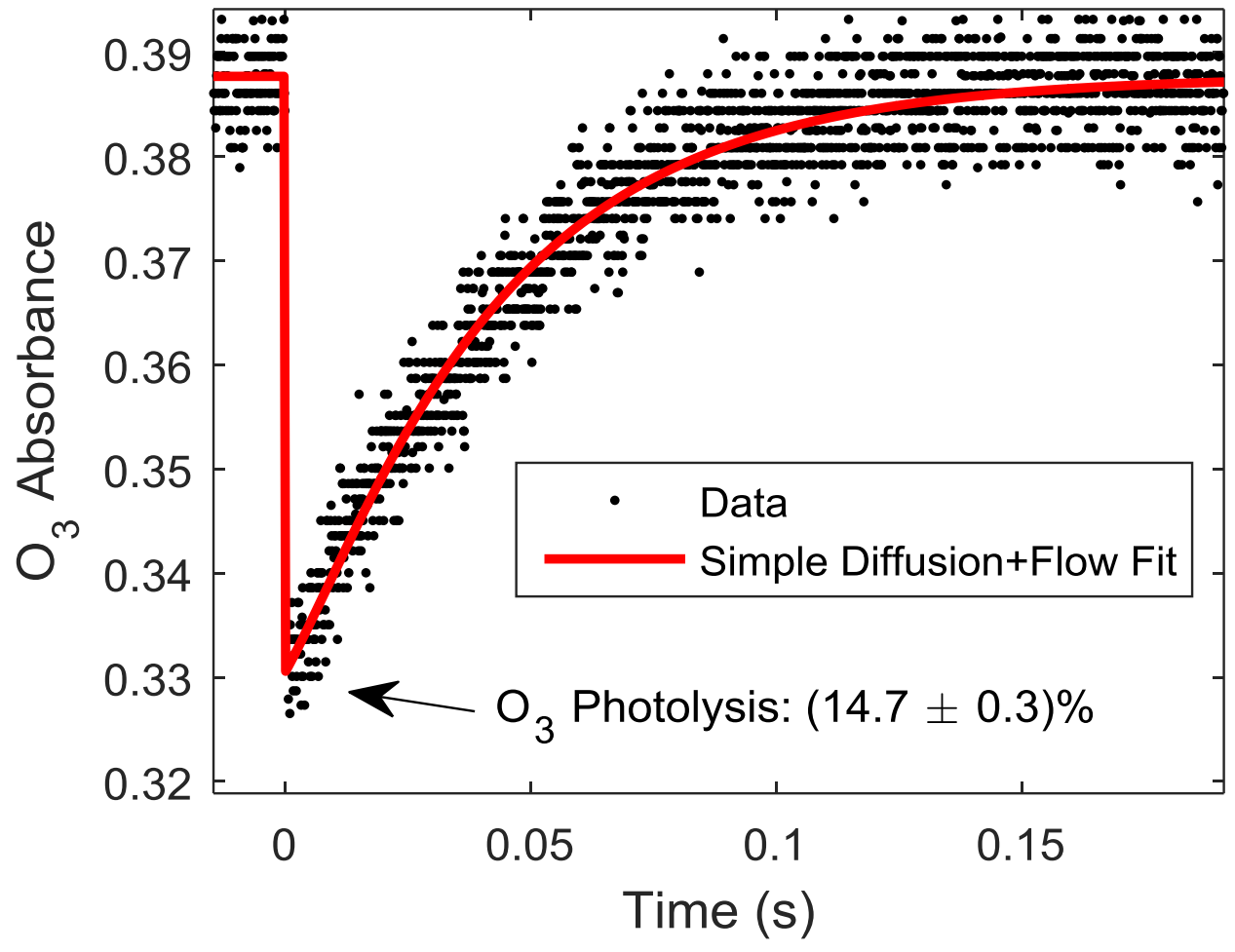

**Fig. S1**. Measurement of Ozone photolysis fraction.

***Experimental conditions:*** Table S1 shows the measurement conditions used for determining $k_{1a}$. The initial concentrations of $D_2$, CO, $N_2$, and $O_3$ were determined by calibrated flow controllers and a capacitance manometer as

$$[N_2]_0 = P\left(\frac{Q_{N2}}{Q_{D2}+Q_{CO}+Q_{N2}+Q_{O3}}\right)$$

where P is the total pressure of the reaction cell and the $Q_X$ are the flows of each gas into the reaction cell. By swapping flow controllers and pressure meters, the concentration uncertainty in this method is estimated to be 7%.

The initial concentrations of $O_3$, $O(^1D)$, and $O_2$ just after photolysis by the YAG beam are given by

$$[O_3]_0 = (1-f_{phot})[O_3]_{LED}$$
$$[O(^1D)]_0 = q_{O(1D)}(f_{phot})[O_3]_{LED}$$
$$[O(^3P)]_0 = q_{O(3P)}(f_{phot})[O_3]_{LED}$$

where $q_{O(1D)} = 0.90$ and $q_{O(3P)} = 0.10$ are the quantum yields (*S1*) of the photolysis of $O_3$ into $O(^1D)$ and $O(^3P)$, respectively, $f_{phot} = 0.15 \pm 0.02$ the $O_3$ photolysis fraction, and $[O_3]_{LED}$ is the steady-state concentration of $O_3$ recorded by the absorption of the 270 nm UV LED.

**Table S1:** Measurement conditions used for the determination of $k_{1a}$. The units for [CO], $[N_2]$, $[D_2]$, and $[O_3]$ are molecules $cm^{-3}$.

| | [CO] | $[N_2]$ | $[D_2]$ | $[O_3]$ | Int. Time (μs) |
|---|---|---|---|---|---|
| **CO Scan** | $7.41 \times 10^{16}$ | $8.89 \times 10^{17}$ | $2.96 \times 10^{17}$ | $1.00 \times 10^{15}$ | 50 |
| | $1.48 \times 10^{17}$ | $8.89 \times 10^{17}$ | $7.41 \times 10^{16}$ | $1.00 \times 10^{15}$ | 50 |
| | $2.96 \times 10^{17}$ | $8.89 \times 10^{17}$ | $7.41 \times 10^{16}$ | $1.00 \times 10^{15}$ | 10 |
| | $2.96 \times 10^{17}$ | $8.89 \times 10^{17}$ | $7.41 \times 10^{16}$ | $1.00 \times 10^{15}$ | 50 |
| | $5.92 \times 10^{17}$ | $8.89 \times 10^{17}$ | $7.41 \times 10^{16}$ | $1.55 \times 10^{15}$ | 10 |
| | $5.92 \times 10^{17}$ | $8.89 \times 10^{17}$ | $7.41 \times 10^{16}$ | $1.00 \times 10^{15}$ | 50 |
| | $5.92 \times 10^{17}$ | $8.89 \times 10^{17}$ | $7.41 \times 10^{16}$ | $1.00 \times 10^{15}$ | 50 |
| | $7.41 \times 10^{17}$ | $8.89 \times 10^{17}$ | $7.41 \times 10^{16}$ | $1.00 \times 10^{15}$ | 10 |
| | $7.41 \times 10^{17}$ | $8.89 \times 10^{17}$ | $7.41 \times 10^{16}$ | $1.00 \times 10^{15}$ | 50 |
| | $8.89 \times 10^{17}$ | $8.89 \times 10^{17}$ | $7.41 \times 10^{16}$ | $1.00 \times 10^{15}$ | 10 |
| | $9.63 \times 10^{17}$ | $8.89 \times 10^{17}$ | $7.41 \times 10^{16}$ | $1.00 \times 10^{15}$ | 10 |
| | $9.63 \times 10^{17}$ | $8.89 \times 10^{17}$ | $7.41 \times 10^{16}$ | $1.00 \times 10^{15}$ | 50 |
| | $1.18 \times 10^{18}$ | $8.89 \times 10^{17}$ | $7.41 \times 10^{16}$ | $1.00 \times 10^{15}$ | 10 |
| | $1.18 \times 10^{18}$ | $8.89 \times 10^{17}$ | $7.41 \times 10^{16}$ | $1.00 \times 10^{15}$ | 50 |
| | $1.18 \times 10^{18}$ | $8.89 \times 10^{17}$ | $7.41 \times 10^{16}$ | $1.00 \times 10^{15}$ | 50 |
| | $1.48 \times 10^{18}$ | $8.89 \times 10^{17}$ | $7.41 \times 10^{16}$ | $1.00 \times 10^{15}$ | 50 |
| **$N_2$ Scan** | $5.92 \times 10^{17}$ | $1.48 \times 10^{17}$ | $7.41 \times 10^{16}$ | $1.00 \times 10^{15}$ | 50 |
| | $5.92 \times 10^{17}$ | $2.96 \times 10^{17}$ | $7.41 \times 10^{16}$ | $1.00 \times 10^{15}$ | 50 |
| | $5.92 \times 10^{17}$ | $5.92 \times 10^{17}$ | $7.41 \times 10^{16}$ | $1.00 \times 10^{15}$ | 50 |
| | $5.92 \times 10^{17}$ | $7.41 \times 10^{17}$ | $7.41 \times 10^{16}$ | $1.00 \times 10^{15}$ | 50 |
| | $5.92 \times 10^{17}$ | $9.92 \times 10^{17}$ | $7.41 \times 10^{16}$ | $1.00 \times 10^{15}$ | 50 |
| | $5.92 \times 10^{17}$ | $1.18 \times 10^{18}$ | $7.41 \times 10^{16}$ | $1.00 \times 10^{15}$ | 50 |
| | $5.92 \times 10^{17}$ | $1.33 \times 10^{18}$ | $7.41 \times 10^{16}$ | $1.00 \times 10^{15}$ | 50 |
| | $5.92 \times 10^{17}$ | $1.33 \times 10^{18}$ | $7.41 \times 10^{16}$ | $1.00 \times 10^{15}$ | 50 |
| | $5.92 \times 10^{17}$ | $1.48 \times 10^{18}$ | $7.41 \times 10^{16}$ | $1.00 \times 10^{15}$ | 50 |
| | $5.92 \times 10^{17}$ | $1.48 \times 10^{18}$ | $7.41 \times 10^{16}$ | $1.00 \times 10^{15}$ | 50 |
| **$O_3$ Scan** | $1.48 \times 10^{17}$ | $8.89 \times 10^{17}$ | $7.41 \times 10^{16}$ | $5.00 \times 10^{14}$ | 50 |
| | $1.48 \times 10^{17}$ | $8.89 \times 10^{17}$ | $7.41 \times 10^{16}$ | $1.00 \times 10^{15}$ | 50 |
| | $1.48 \times 10^{17}$ | $8.89 \times 10^{17}$ | $7.41 \times 10^{16}$ | $2.50 \times 10^{15}$ | 50 |
| | $1.48 \times 10^{17}$ | $8.89 \times 10^{17}$ | $7.41 \times 10^{16}$ | $5.00 \times 10^{15}$ | 50 |
| | $2.96 \times 10^{17}$ | $8.89 \times 10^{17}$ | $9.57 \times 10^{16}$ | $1.00 \times 10^{14}$ | 50 |
| | $2.96 \times 10^{17}$ | $8.89 \times 10^{17}$ | $8.74 \times 10^{16}$ | $5.00 \times 10^{14}$ | 50 |
| | $2.96 \times 10^{17}$ | $8.89 \times 10^{17}$ | $7.41 \times 10^{16}$ | $1.00 \times 10^{15}$ | 50 |
| | $2.96 \times 10^{17}$ | $8.89 \times 10^{17}$ | $7.41 \times 10^{16}$ | $2.00 \times 10^{15}$ | 50 |
| | $2.96 \times 10^{17}$ | $8.89 \times 10^{17}$ | $8.15 \times 10^{16}$ | $5.00 \times 10^{15}$ | 50 |
| **$D_2$ Scan** | $2.96 \times 10^{17}$ | $8.89 \times 10^{17}$ | $1.48 \times 10^{17}$ | $1.00 \times 10^{15}$ | 50 |
| | $2.96 \times 10^{17}$ | $8.89 \times 10^{17}$ | $2.96 \times 10^{17}$ | $1.00 \times 10^{15}$ | 50 |
| | $2.96 \times 10^{17}$ | $8.89 \times 10^{17}$ | $5.92 \times 10^{17}$ | $1.00 \times 10^{15}$ | 50 |
| | $2.96 \times 10^{17}$ | $8.89 \times 10^{17}$ | $8.89 \times 10^{17}$ | $1.00 \times 10^{15}$ | 50 |
| | $2.96 \times 10^{17}$ | $8.89 \times 10^{17}$ | $1.18 \times 10^{18}$ | $1.00 \times 10^{15}$ | 50 |
| **OD(v=1) lifetime** | 0 | $9.98 \times 10^{17}$ | $8.32 \times 10^{16}$ | $1.00 \times 10^{16}$ | 50 |
| | $2.06 \times 10^{16}$ | $9.91 \times 10^{17}$ | $8.26 \times 10^{16}$ | $1.00 \times 10^{16}$ | 50 |
| | $4.08 \times 10^{16}$ | $9.79 \times 10^{17}$ | $8.16 \times 10^{16}$ | $1.00 \times 10^{16}$ | 50 |
| | $5.64 \times 10^{16}$ | $9.67 \times 10^{17}$ | $8.06 \times 10^{16}$ | $1.00 \times 10^{16}$ | 50 |
| | $8.19 \times 10^{16}$ | $9.83 \times 10^{17}$ | $8.19 \times 10^{16}$ | $1.00 \times 10^{16}$ | 50 |
| | $9.79 \times 10^{16}$ | $9.80 \times 10^{17}$ | $8.17 \times 10^{16}$ | $1.00 \times 10^{16}$ | 50 |

§2. Data extraction & analysis

***Spectral acquisition:*** The transmitted mid-IR light was spatially dispersed using a Virtually Imaged Phased Array (VIPA) and detected using a FLIR SC6000 InSb camera, in the same manner as Fleisher *et al.* (*S2*). The camera integration (50 μs or 10 μs integration time) was synchronized to the Nd:YAG photolysis pulse. A digital delay generator sets the variable delay times from the photolysis pulse. Since we are not resolving individual frequency comb modes, we calibrated our frequency axis each day to known $D_2O$ line positions from Ref. (*S3*).

The experiment was conducted at a 10 Hz repetition rate, set by the maximum repetition rate of the pulsed Nd:YAG laser. A "reference" image was collected directly prior to each Nd:YAG pulse and "signal" images were collected at various delay times following the Nd:YAG pulse. Since the InSb camera has a dark current offset that drifts with ambient temperature, a "background" image was also collected at the same repetition rate by briefly blocking the camera with an optical shutter.

After collecting each set of images, the absorbance was constructed in the following manner

$$A = -\log\left(\frac{S-B}{R-B}\right),$$

where *S*, *R*, and *B* are the signal, reference, and background images, respectively. Due to slowly-varying baseline fluctuations in the transmission through the cavity, a sliding average was subtracted from the measured absorbance as a function of wavenumber, forming a "high-passed" signal $\tilde{A} = A - H[A]$, where *H* is the sliding average function. Following the "high-pass" filter operation, the error at each point in the spectrum is estimated by taking the standard deviation of the surrounding points. In this manner, each collected spectrum is assigned a value and error corresponding to $\tilde{A} \pm \delta\tilde{A}$. Averaging many of these values yields an average value $\tilde{A}_{mean} \pm \delta\tilde{A}_{mean}$. Since this "sliding standard deviation" operation includes some of the absorption peaks in the spectrum, it is a slight overestimate of the error in the spectrum.

***Spectral fitting:*** In direct absorption spectroscopy, the concentration of a species is related to the transmission of a probe beam through the relation

$$\frac{I_S(\tilde{\nu})}{I_R(\tilde{\nu})} = e^{-n\sigma(\tilde{\nu})l},$$

where $I_S(\tilde{\nu})$ and $I_R(\tilde{\nu})$ are the light intensities with and without the sample, $n$ is the molecular concentration in molecules cm$^{-3}$, $\sigma(\tilde{\nu})$ is the molecular absorption cross section in cm$^2$, $l$ is the path length through the sample in cm, and $\tilde{\nu}$ is wavenumber in cm$^{-1}$. For this experiment, $I_R(\tilde{\nu})$ is recorded 4 ms before the photolysis pulse and $I_S(\tilde{\nu})$ is recorded after the photolysis pulse by the InSb camera. If multiple species are present, the transmission versus time is now given by

$$-\frac{1}{l}\log\left(\frac{I_S(\tilde{\nu})}{I_R(\tilde{\nu})}\right) = n_{\mathrm{A}}(t)\sigma_{\mathrm{A}}(\tilde{\nu}) + n_B(t)\sigma_B(\tilde{\nu}) + \ldots,$$

where A and B are two sample molecules. If $\sigma_{\mathrm{A}}(\tilde{\nu})$, $\sigma_B(\tilde{\nu})$ are linearly independent as a function of wavelength, then $n_{\mathrm{A}}(t)$, $n_{\mathrm{B}}(t)$ are determined uniquely through linear regression. $\sigma(\tilde{\nu})$ is related to the molecular line strength $S$ through $\sigma(\tilde{\nu}) = Sg(\tilde{\nu} - \tilde{\nu}_0)$ where $g(\tilde{\nu} - \tilde{\nu}_0)$ is the area normalized lineshape function. In our case, $g(\tilde{\nu} - \tilde{\nu}_0)$ is a Gaussian function with FWHM = 900 MHz. This is significantly larger than the molecular Doppler width, so convolution with the molecular lineshape is neglected.

Since we "high-pass" the measured absorbance, $A(\tilde{\nu})$, to reduce the effects of cavity fluctuations, it is also necessary to perform the same operation on the calculated molecular cross sections. This will not affect the fitted concentration values, since the sliding average operation $H[A]$ is a linear function and thus

$$-\frac{1}{l}\log\left(A - H[A]\right) = n_{\mathrm{A}}(t)(\sigma_{\mathrm{A}}(\tilde{\nu}) - \mathrm{H}[\sigma_{\mathrm{A}}(\tilde{\nu})]) + n_{\mathrm{B}}(t)(\sigma_{\mathrm{B}}(\tilde{\nu}) - \mathrm{H}[\sigma_{\mathrm{B}}(\tilde{\nu})]) + \ldots,$$

where A is the absorbance, given by $A = -\log\left(\frac{I_S(\tilde{\nu},t)}{I_R(\tilde{\nu})}\right)$.

***Spectral line intensities:*** Line positions for $D_2O$ were obtained from Ref. (*S3*). Line strengths were measured by Dr. Robert A. Toth at Caltech/JPL, and generously provided through private communication with Dr. Keeyoon Sung of JPL. Line positions for OD(v=0,1) were obtained from Ref. (*S4*). PGopher (*S5*) was used along with fit parameters from Ref. (*S4*) to obtain relative line strengths for each line in the spectrum. The OD v=0 and v=1 transition dipole moments $|\mu^{OD}{}_{0\rightarrow1}|$ = 0.0303 and $|\mu^{OD}{}_{1\rightarrow2}|$ = 0.0386 D were obtained from mass-scaling the OH transition dipole moments. The $|\mu^{OH}{}_{0\rightarrow1}|$ = 0.0343 and $|\mu^{OH}{}_{1\rightarrow2}|$ = 0.0408 transition dipole moments were calculated using the RKR potential and dipole moment functions reported by Nesbitt and coworkers (*S6, 7*). The error in the transition dipole moments is estimated to be <10% for OD. *trans*-DOCO $\nu_1$ ro-vibrational parameters were obtained from Ref. (*S8*) and used to simulate the ro-vibrational spectrum in PGopher. As there are no known measurements of the *trans*-DOCO band intensity, we assume a *trans*-DOCO $\nu_1$ band strength of $S_{trans\text{-DOCO}}$ = 65 ± 5 km/mol, estimated from a series of anharmonic VPT2 vibrational calculations performed at the CCSD(T)/ANOn (n = 0,1,2) and CCSD(T)/cc-pCVXZ (X = D,T,Q) levels of theory (personal communication with J. F. Stanton).

***Photolysis Path Length and Finesse:*** In cavity-enhanced spectroscopy, the path length $l$ is given by the physical path length multiplied by a factor of $\beta F/\pi$, where $F$ is the finesse of the optical cavity and $1 \leq \beta \leq 2$ is a parameter that arises when a sweep-lock is used(*S9*). In addition, the path length is reduced to the width of the photolysis beam, $l_{\text{phot}}$. Thus, the effective optical absorption path length is given by

$$l_{eff} = \frac{\beta F l_{phot}}{\pi}.$$

The finesse of the cavity as a function of wavelength was measured using cavity ringdown and is shown in Fig. S2.

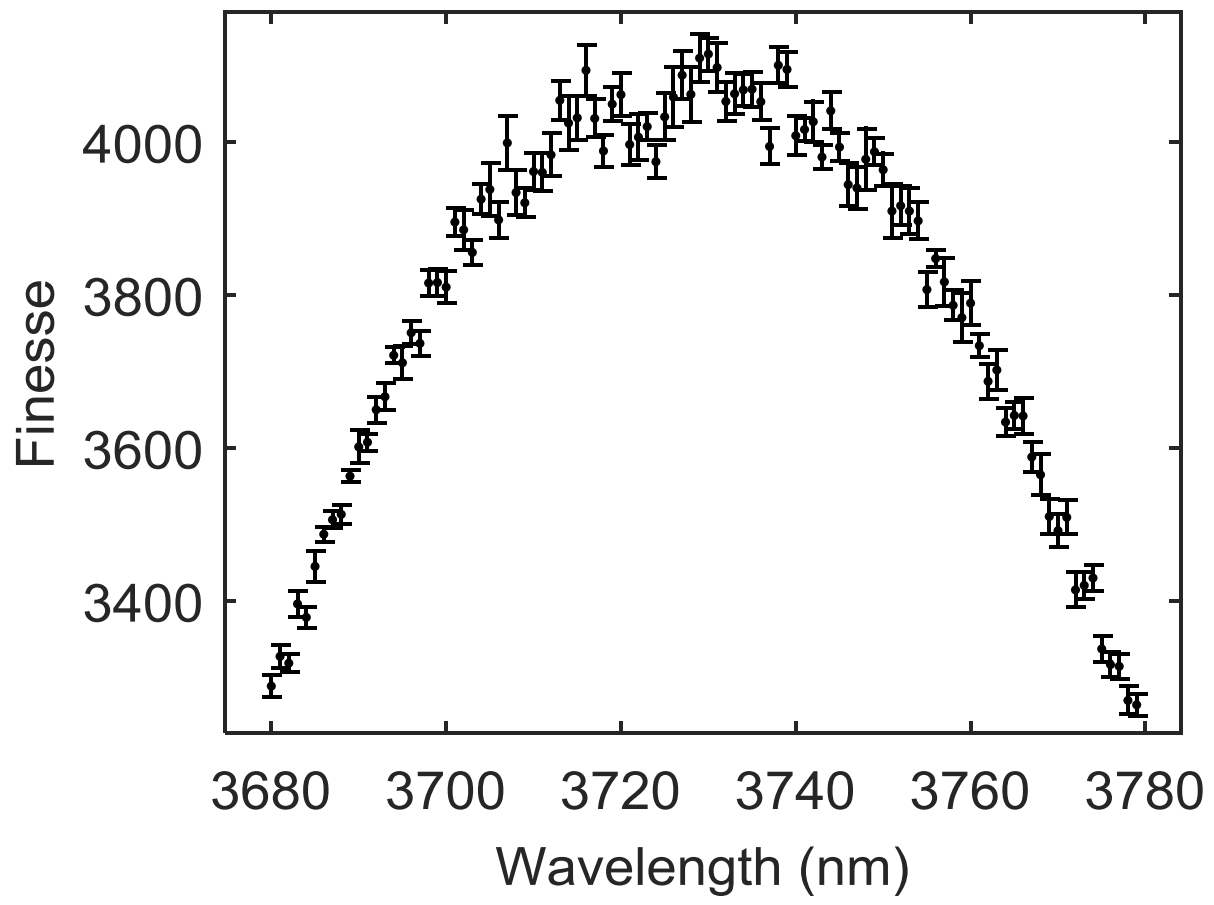

**Fig. S2**. Finesse from spectrally resolved cavity-ringdown measurements.

The photolysis path length, $l_{phot}$, was determined in two ways: 1) The width (46 ± 5 mm) of the burn spot on a photographic film from YAG beam and 2) a razor blade scan across the beam and fitting the OD concentration at each point of the scan (Fig. S3). The razor blade method gave a beam width of 42 ± 4 mm, which is in agreement with the photographic film method. The weighted average of these two methods is 44 ± 3 mm.

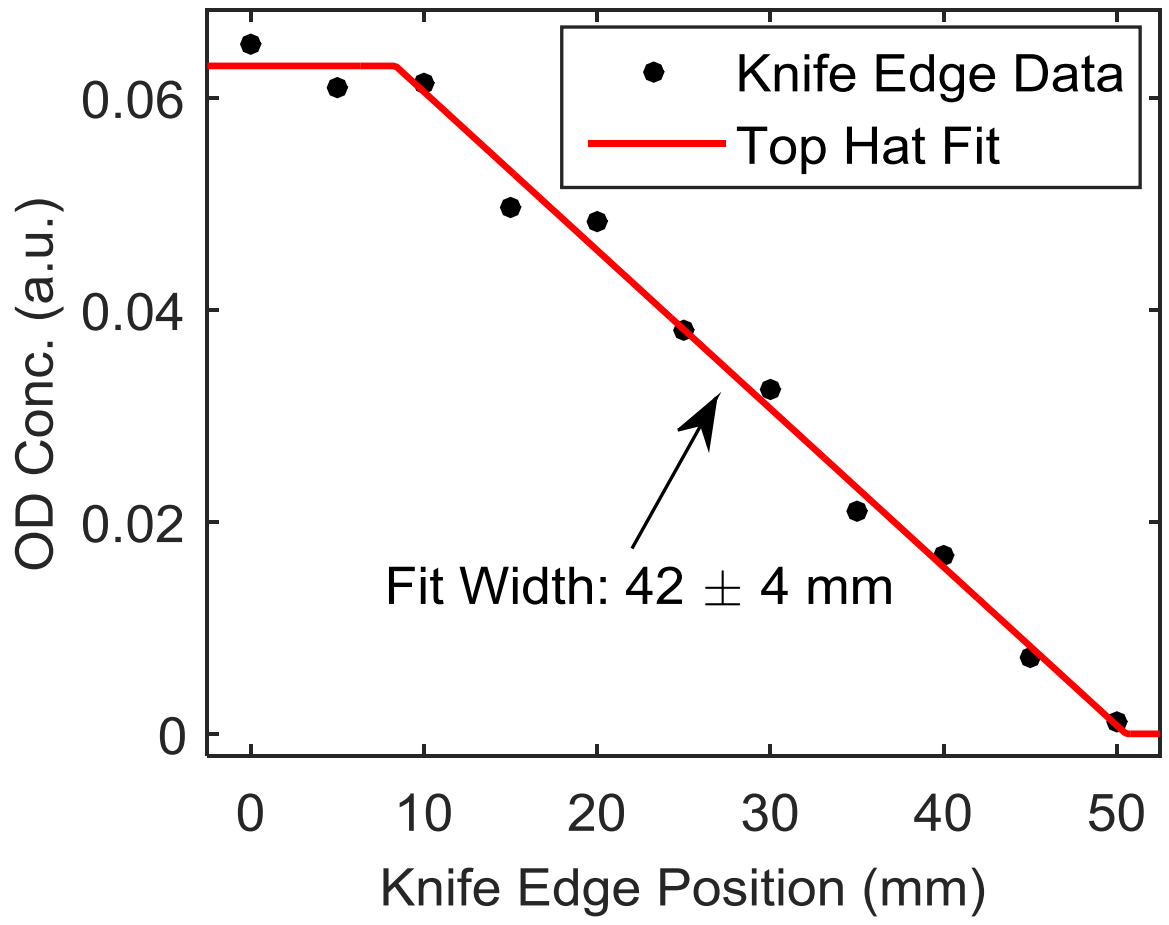

**Fig. S3**. Knife edge scan of YAG beam.

The error in the effective path length is given as

$$\delta l_{eff} = l_{eff}\sqrt{\left(\frac{\delta\beta}{\beta}\right)^2 + \left(\frac{\delta F}{F}\right)^2 + \left(\frac{\delta l_{phot}}{l_{phot}}\right)^2},$$

which yields $l_{eff}$ = 58±4 m at the finesse peak of 3725 nm.

## §3. Sources of error

***Spectral interference from $D_2O$:*** In performing a linear spectral fit to OD, $D_2O$, and *trans*-DOCO, it is possible that absorption from one species may interfere with another. This cross-contamination effect can be exacerbated if the lineshape of the fit does not exactly match the experiment. The largest cross-contamination effect in our experiment is between OD and $D_2O$, since half of the OD lines in our spectral window are contaminated by strong $D_2O$ transitions. However, this is nearly negligible in the first 100 μs, where $D_2O$ concentrations are low. Based on a comparison of contaminated and uncontaminated OD absorption features, we estimate that the systematic error due to cross-contamination is <1%.

***Initial Rate Method:*** At early times, we expect DOCO to behave according to the first-order differential equation,

$$[\mathrm{DOCO}](t) = k_{1a}[\mathrm{CO}][\mathrm{OD}](t) - k_{loss}[\mathrm{X}] \;,$$

where [X] is the primary loss partner for DOCO and [OD]($t$) refers to the time-dependent concentration of OD in the ground vibrational state, OD(v=0). With the initial condition that [DOCO(t=0)] = 0, we can solve this equation directly for [DOCO](t) in terms of [OD](t), which yields

$$[\mathrm{DOCO}](t) = k_{1\mathrm{a}}[\mathrm{CO}]\int_0^t e^{-(k_{\mathrm{loss}}[\mathrm{X}])(t-u)}[\mathrm{OD}](u)du \;.$$

To obtain an analytic form for OD(t), we fit a sum of exponentials to our experimental data, constrained by [OD]($t$=0) = 0.

$$[\mathrm{OD}](t) = a_1 e^{-b_1 t} + a_2 e^{-b_2 t} - (a_1 + a_2)e^{-b_3 t} \;.$$

Here, $b_1$ and $b_2$ are bi-exponential decay terms while $b_3$ is a rise term. The $b_3$ rise term for OD(v=0) is directly related to the decay of OD(v=1) where it originates. OD(v=1) decay will be discussed in more detail in the following sections. Using this expression for OD($t$), DOCO($t$) is given by:

$$[\mathrm{DOCO}](t) = k_{1a}[\mathrm{CO}]\left( a_1 \frac{e^{-b_1 t} - e^{-r_{loss} t}}{b_1 - r_{loss}} + a_1 \frac{e^{-b_2 t} - e^{-r_{loss} t}}{b_2 - r_{loss}} - \left(a_1 + a_2\right) \frac{e^{-b_3 t} - e^{-r_{loss} t}}{b_3 - r_{loss}} \right).$$

DOCO($t$) contains two free parameters in this expression, $k_{1\mathrm{a}}$, and $r_{\mathrm{loss}}$ ( ≡ $k_{\mathrm{loss}}$[X]). We fit a $r_{\mathrm{loss,exp}} = (4.0\pm0.4)\times10^4\ \mathrm{s}^{-1}$ to all data with constant [$O_3$], while our fit value of $k_{1\mathrm{a}}$ varies with $N_2$ and CO.

From our fit values of the bimolecular rate constant $k_{1a}$, we determine the termolecular rates $k_{1a}^{(CO)}$ and $k_{1a}^{(N2)}$ from a multidimensional linear regression to the expression

$$k_{1a} = k_{1a}^{(\mathrm{CO})}[\mathrm{CO}] + k_{1a}^{(\mathrm{N2})}[\mathrm{N}_2].$$

The statistical error in our values of $k_{1a}^{(CO)}$ and $k_{1a}^{(N2)}$ are established from the variation in the fit residuals.

***Uncertainty in the Initial Rate Method:*** To investigate the variation in our fitted $k_{1a}$ values with integration time, we divided our values for $k_{1a}$ (see main text) for 50 and 10 μs and plotted these values as a function of CO concentration. We display this value, $r = \frac{k_{1a,10\mu s}}{k_{1a,50\mu s}}$, with CO in Fig. S4.

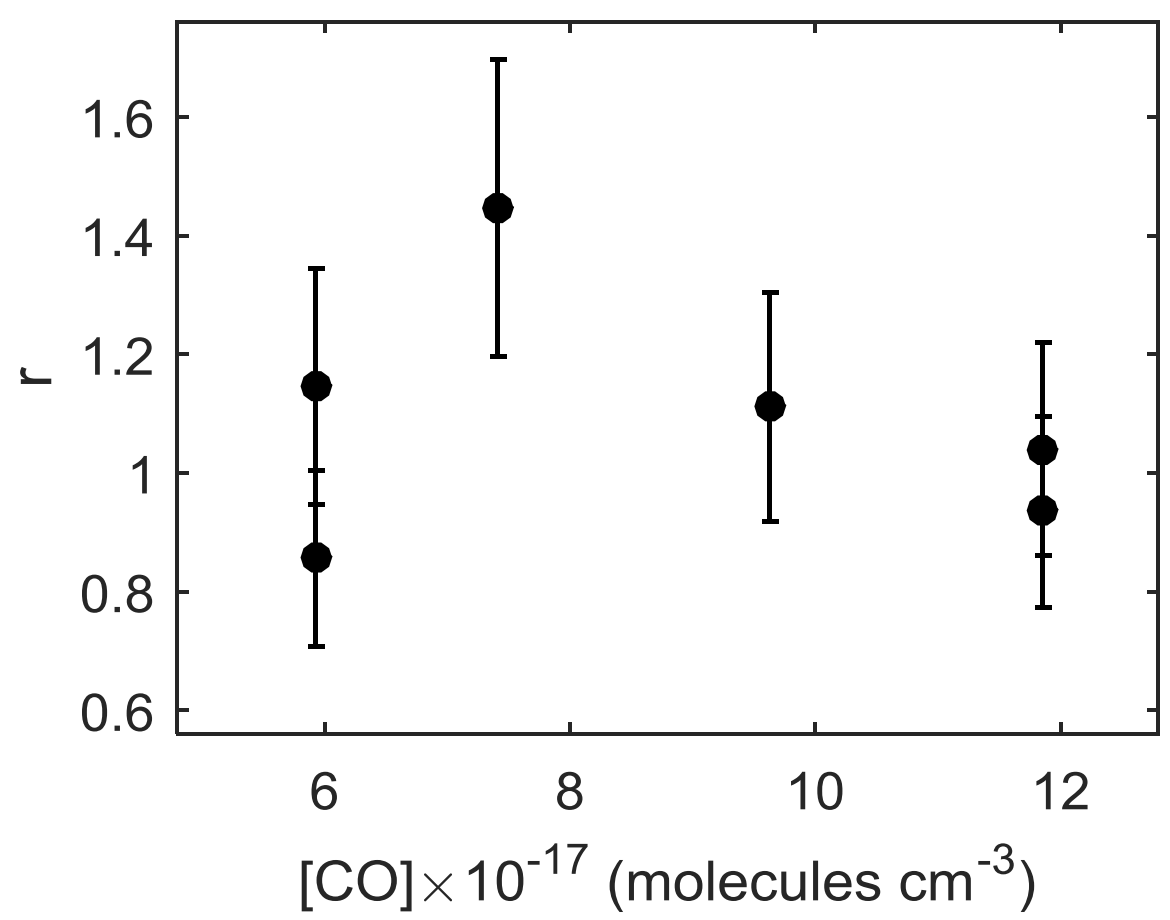

**Fig. S4**. The ratio of retrieved $k_{1a}$ values for 50 μs and 10 μs integration times.

In order to estimate a general systematic error in a given data set, we calculate the weighted mean and standard deviation of this data, which are $\bar{r} = 1.09$ and $\sigma_r = 0.21$, respectively. We interpret $\bar{r} - 1 = 9\%$ as a systematic shift due to our 50 μs integration time and $\sigma_r$ as an estimate of the statistical variation of $k_{1a}$ with respect to integration time and [CO].

***Effect of OD Vibrational Excitation:*** Since vibrationally hot OD(v>0) was generated under our experimental conditions, we investigated the effect of vibrational quenching of hot OD under conditions relevant for fitting the OD + CO → DOCO rate. First, $O(^1D) + D_2 \rightarrow OD + D$ promptly produces excited OD up to the 4$^{th}$ vibrational state (*S10*). The second reaction, $D + O_3 \rightarrow OD + O_2$, continually generates hot OD up to the 9$^{th}$ vibrational state (*S11*).

Due to the broad bandwidth and high sensitivity of our kinetics apparatus, we were simultaneously able to detect OD(v=0) and OD(v=1) in a time-resolved manner during each experimental run. Fig. S5 show an acquired spectrum containing both strong OD(v=0) (blue) and OD(v=1) (red) transitions.

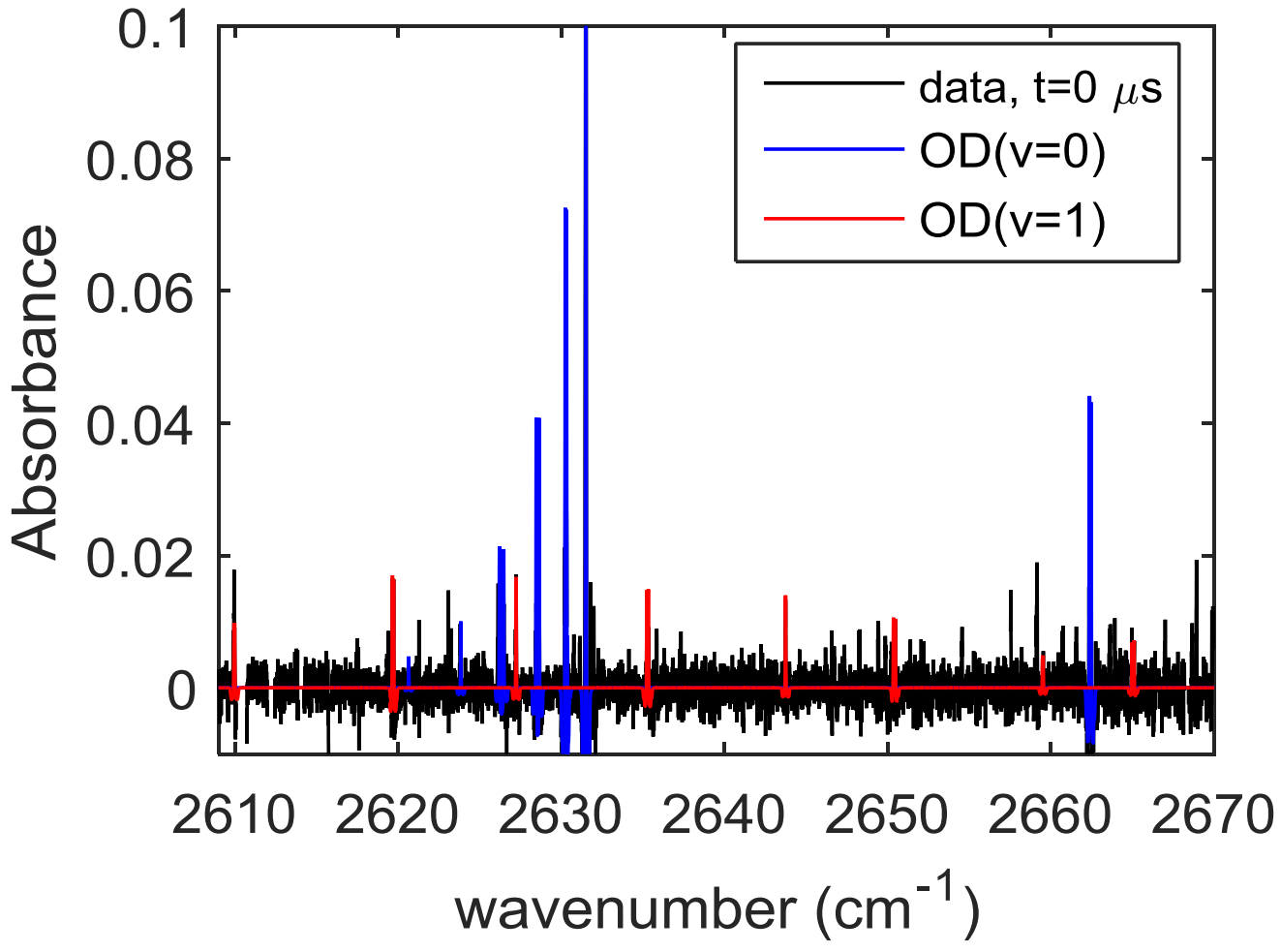

**Fig. S5.** Representative spectrum of OD(v=0) and OD(v=1) for measuring [OD(v>0)].

To constrain the extent to which OD(v>0) introduces error into our overall determination of $k_{1a}$, we conducted experiments to measure both the density and lifetime of OD(v=1) over a range of CO densities. Experiments were conducted with $[O_3] = 1\times10^{16}$ molecules $cm^{-3}$, which provided an upper limit for the vibrationally excited OD population. [CO] ranged from $0\text{-}9\times10^{16}$ molecules $cm^{-3}$, which is well below the lowest [CO] used in our $k_{1a}$ measurement. We observed high signal-to-noise ratio OD(v=1) transitions, but not any OD(v>1), indicating that either 1) the densities are too low and/or 2) the lifetimes are too short for higher exited states. The results from this experiment are shown in Fig S6. At our operating conditions for determining $k_{1a}$, the lifetime of OD(v=1) is <5 μs. The measured densities for OD(v=1) are also <10% of OD(v=0). Therefore, even if OD(v>0) forms ground state *trans*-DOCO at the same rate as OD(v=0), this would only introduce a <10% uncertainty in our measurement, which has been included in our systematic error budget (Table S3).

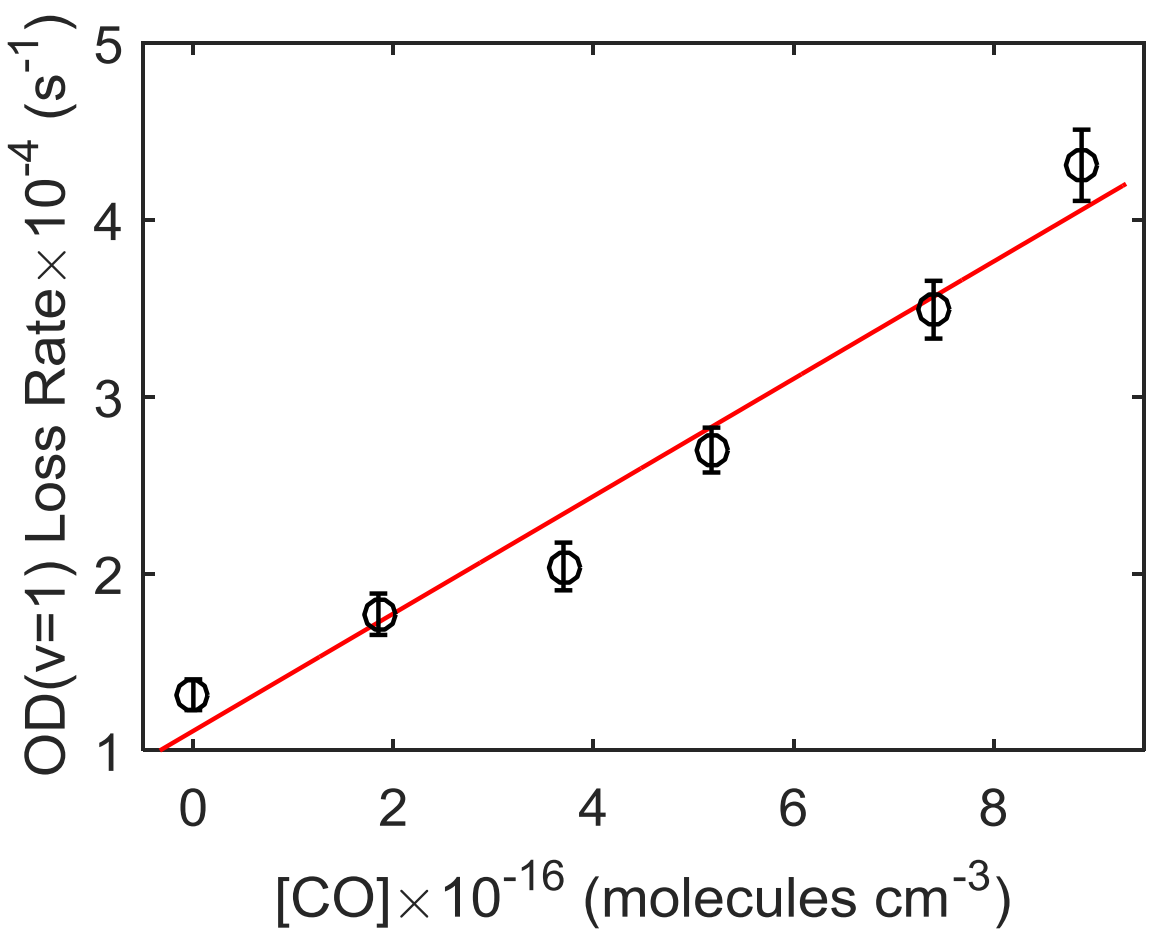

**Fig. S6.** First order decay rate of OD(v=1) as a function of [CO]. The fitted rate constant for OD(v=1) loss is $(3.3\pm0.2)\times10^{-13}$ $cm^3$ $molecules^{-1}$ $s^{-1}$.

***Effect of $D_2$:*** In order to determine the systematic effects of large [$D_2$], we varied [$D_2$] under constant [$N_2$]=$8.9\times10^{17}$ molecules $cm^{-3}$, [CO] =$3.0\times10^{17}$ molecules $cm^{-3}$, and [$O_3$]=$1.0\times10^{15}$ molecules $cm^{-3}$. The results, shown in Fig. S7, show no statistically significant variation with [$D_2$]. While there are no literature estimates of the DOCO* + $D_2$ → DOCO + $D_2$ quenching efficiency, we might expect this rate to be slower than the $N_2$ rate by about a factor of 2, given similar comparisons in toluene (*S12*). In this case, we would expect the termolecular rate $k_{1a}$ to change by about $0.5\times10^{14}$ $cm^3$ $molecules^{-1}$ $s^{-1}$ with a $1\times10^{18}$ molecules $cm^{-3}$ variation in [$D_2$]. The magnitude of the uncertainty in Fig. S7 is about $0.6\times10^{14}$ $cm^3$ $molecules^{-1}$ $s^{-1}$, and hence a termolecular $D_2$ effect, if it exists, is masked by the noise.

Ideally, $k_{1a}$ is determined in the limit of [$D_2$] → 0. To determine the shift associated with nonzero $D_2$, we performed a linear fit to this data, resulting in a slope of $(1.3 \pm 0.8) \times 10^{-33}$ $cm^6$ $molecules^{-2}$ $s^{-1}$ and an offset of $(1.44 \pm 0.06) \times 10^{-14}$ $cm^3$ $molecules^{-1}$ $s^{-1}$. At our typical operating concentration of [$D_2$] = $7.4\times10^{16}$ molecules $cm^{-3}$, this results in a systematic shift of (0.7 ± 0.4)%. However, this analysis assumes that $k_{1a}$ is linear with $D_2$. As a much more conservative estimate of our systematic error, we use the mean-normalized standard deviation, $\frac{\sigma_{k_{1a}}}{k_{1a}} = 8\%$, of the data in Fig. S7 as the fractional statistical error due to nonzero [$D_2$].

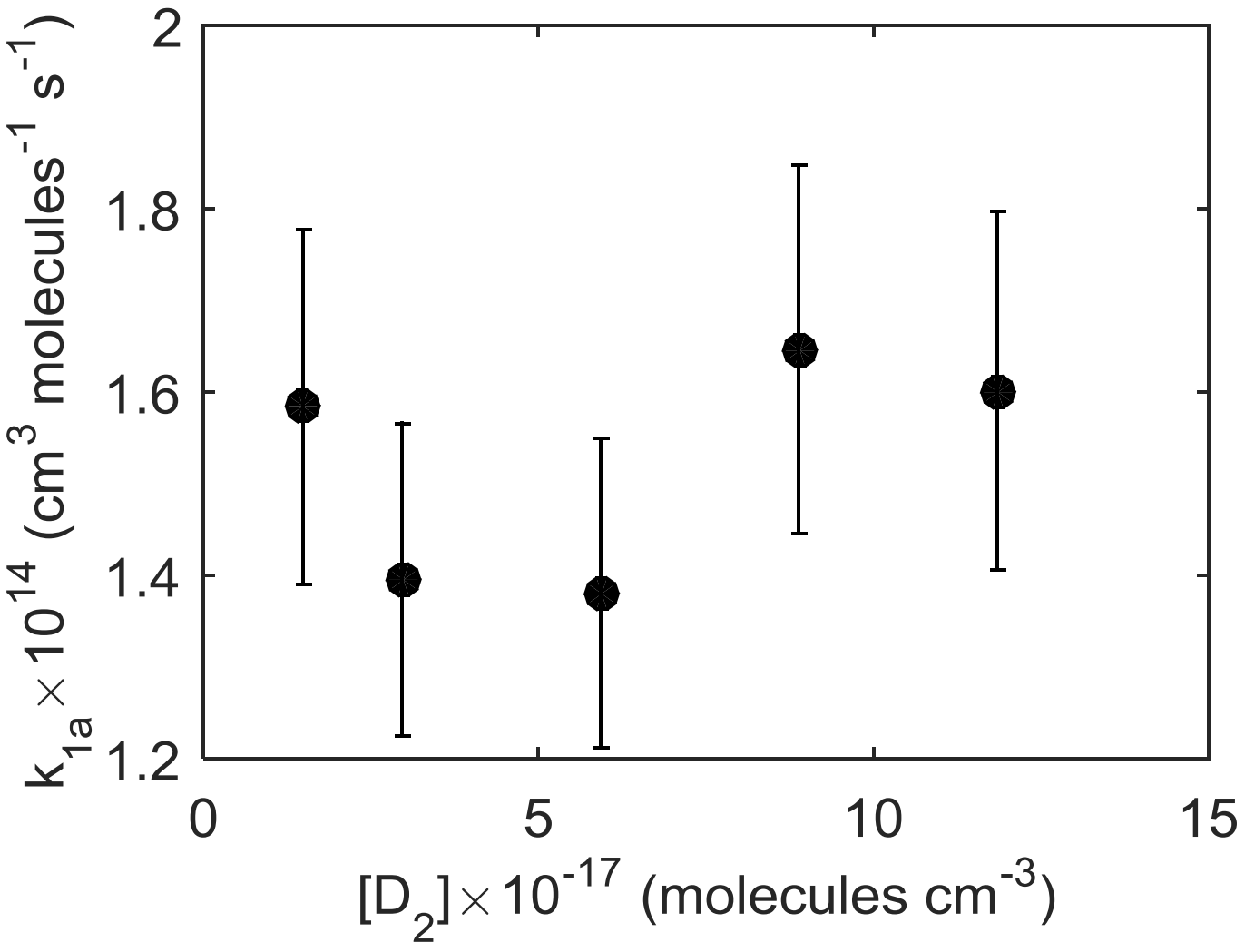

**Fig. S7**. Variation of $k_{1a}$ with $D_2$ concentration.

***Effect of $O_3$:*** In order to determine the systematic effects of $[O_3]$ in our measurement of $k_{1a}^{(CO)}$ and $k_{1a}^{(N2)}$, we varied $[O_3]$ under constant $[N_2]=8.9\times10^{17}$ molecules cm$^{-3}$, $[CO]=1.5\times10^{17}$ molecules cm$^{-3}$, and $[D_2]= 7.4\times10^{16}$ molecules cm$^{-3}$. With fixed $[O_3]=1\times10^{15}$ molecules cm$^{-3}$, we find our experimental data is consistent with a constant $r_{loss,exp} = (4.0\pm0.4)\times10^4$ s$^{-1}$, independent of $[N_2]$, $[D_2]$, and [CO]. In the case of varying $[O_3]$, however, we find that a constant $r_{loss,exp}$ term results in poor fits and also an observed systematic variation of $k_{1a}$ with $[O_3]$. We therefore make the assumption that $r_{loss,exp}$ scales with $[O_3]$, i.e. $r_{loss,exp} = k_{O3}[O_3]$, where $k_{O3} = 4.0\times10^{-11}$ cm$^3$ molecules$^{-1}$ s$^{-1}$ is fixed by the results of the constant $[O_3]$ data. The retrieved values of $k_{1a}$ vs $[O_3]$, shown in Fig. S8, display a weak dependence of $k_{1a}$ on $[O_3]$. The weighted mean and standard deviation of this data are $\bar{k}_{1a} = 1.2\times10^{-14}$ cm$^3$ molecules$^{-1}$ s$^{-1}$ and $\sigma_{k1a} = 1.3\times10^{-15}$ cm$^3$ molecules$^{-1}$ s$^{-1}$, respectively. Ideally, $k_{1a}$ is determined in the limit of $[O_3] \rightarrow 0$. Since we do not see a systematic variation of our retrieved $k_{1a}$ value with $[O_3]$, we interpret $\frac{\sigma_{k_{1a}}}{\bar{k}_{1a}} = 11\%$ as a maximum statistical error due to nonzero $[O_3]$.

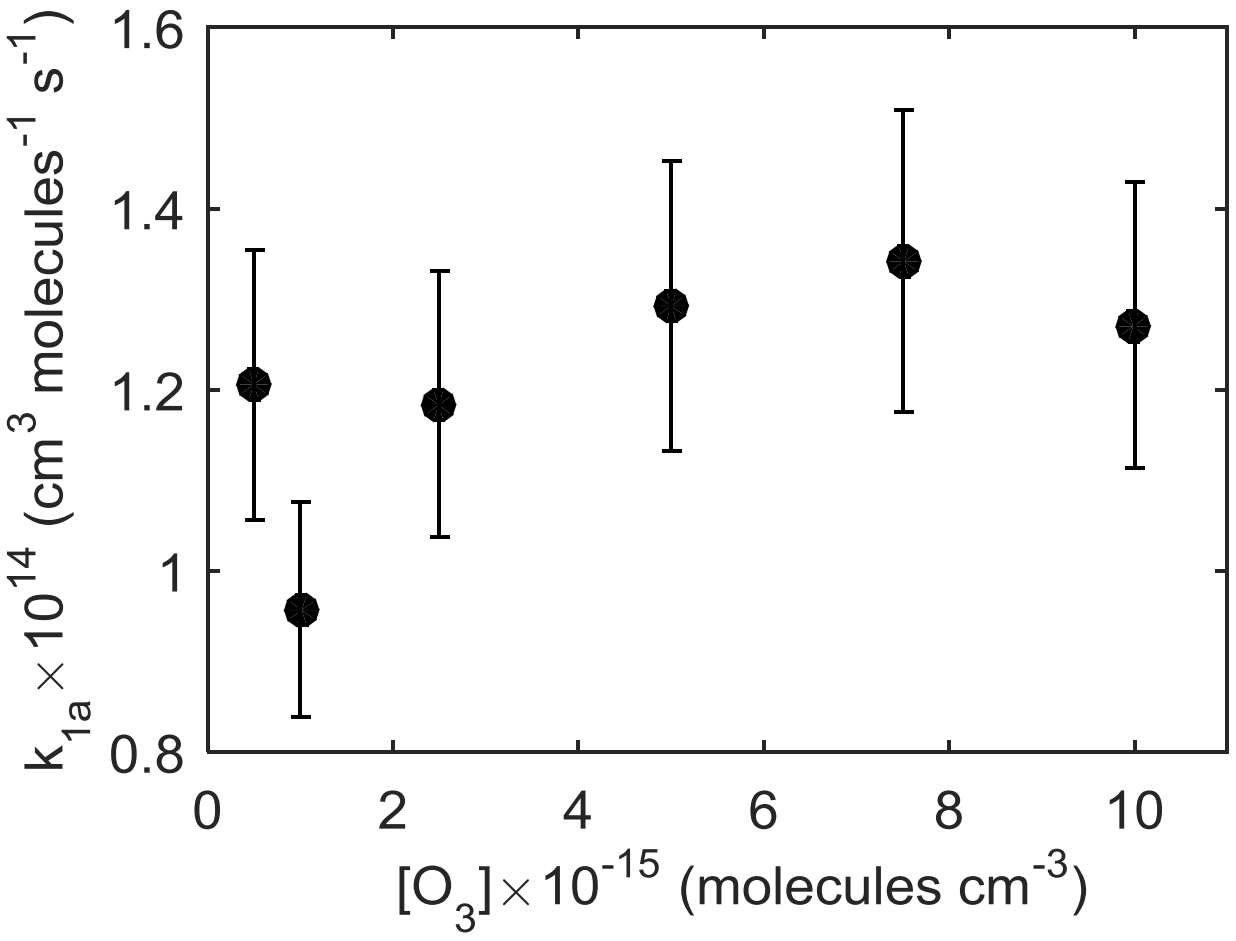

**Fig. S8**. Variation of $k_{1a}$ with $O_3$ concentration.

§4. Rate Equation Model

A full rate equation model that includes all of the most relevant rates for the reaction of OD + CO is given in Table S2. The system of stiff differential equations was integrated using the SimBiology software package from MathWorks and also by the Kintecus software package (*S13*), both of which were in strong agreement. As the experimental data were integrated over 50 or 10 μs, we also boxcar-averaged the results from the rate equation model to fit to experimental data.

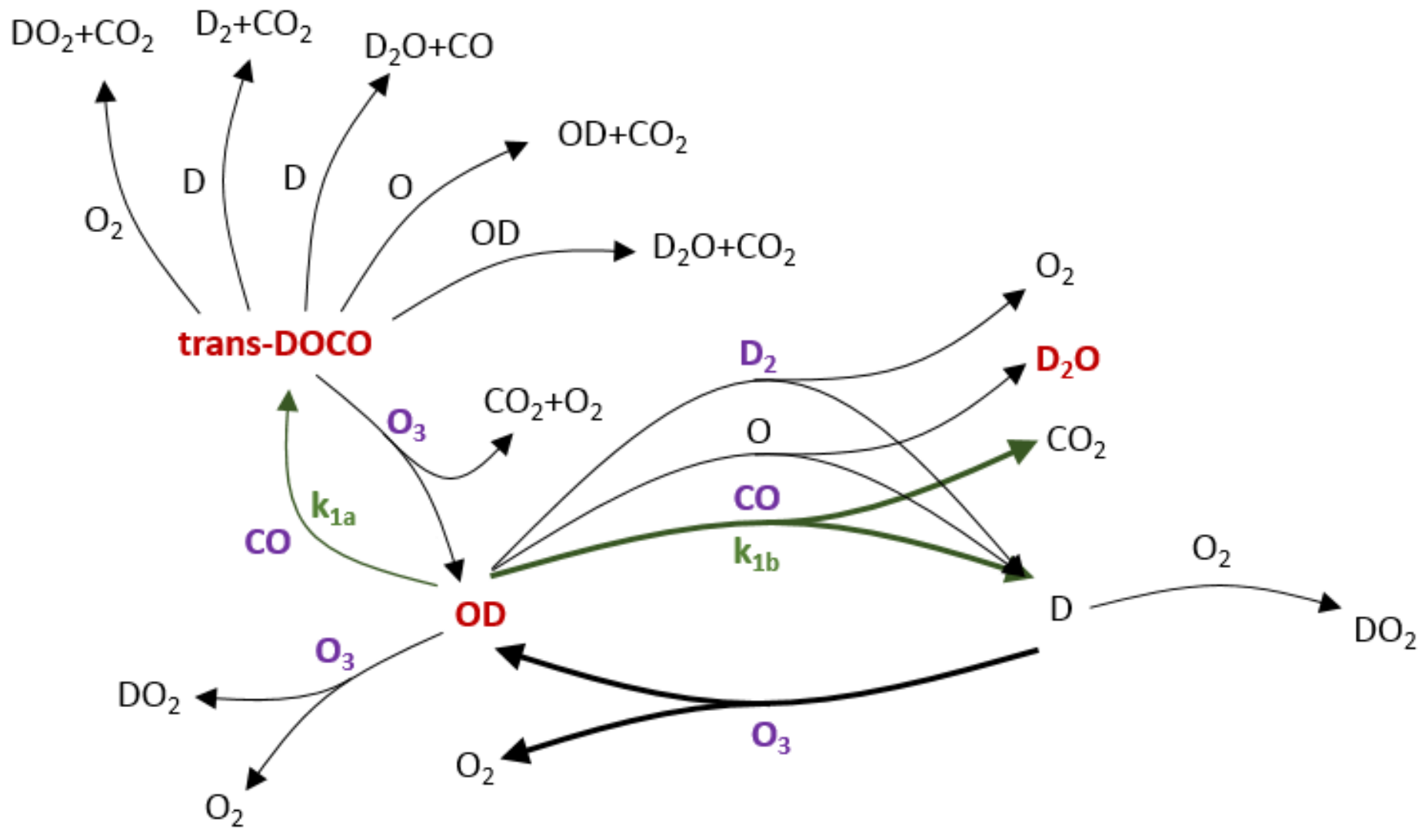

**Fig. S9**. Basic schematic of the rate equation model used in the present studies. Absolute time-dependent concentrations of the red molecules (*trans*-DOCO, OD(v=0), OD(v=1), and $D_2O$) are measured through cavity-enhanced absorption spectroscopy, while the concentrations of the precursors (purple) are fixed by controlling the flows of $N_2$, CO, and $D_2$ and measuring the UV absorption of $O_3$, respectively. The two relevant OD + CO branching reactions are indicated in green.

**Table S2:** Rates used for modelling the OD + CO reaction. Units for termolecular rates are $cm^6$ $molecules^{-2}$ $s^{-1}$ and $cm^3$ $molecules^{-1}$ $s^{-1}$ for $k_0^{300}$ and $k_\infty^{300}$, respectively.

| Category | Reaction | Rate ($cm^3$ $molecules^{-1}$ $s^{-1}$) | Source[1] | $S_{OD}$ (%) | $S_{DOCO}$ (%) | Ref(s) |
|---|---|---|---|---|---|---|
| $O(^1D)$ + X | $O(^1D) + N_2 \rightarrow O + N_2$ | $(3.1 \pm 0.3) \times 10^{-11}$ | DM | 63 | 58 | (*S1*),(*S14*) |
| | $O(^1D) + CO \rightarrow O + CO$ | $(5.8 \pm 1.2) \times 10^{-11}$ | DM | 12 | 11 | (*S15*) |
| | $O(^1D) + O_2 \rightarrow O + O_2$ | $(3.95 \pm 0.4) \times 10^{-11}$ | DM | 0 | 0 | (*S1*), (*S16*) |
| | $O(^1D) + O_3 \rightarrow O_2 + O_2$ | $(1.2 \pm 0.2) \times 10^{-10}$ | DM | 0 | 0 | (*S1*) |
| | $O(^1D) + O_3 \rightarrow O_2 + O + O$ | $(1.2 \pm 0.2) \times 10^{-10}$ | DM | 0 | 0 | (*S1*) |
| | $O(^1D) + D_2 \rightarrow OD + D$ | $(1.1 \pm 0.1) \times 10^{-10}$ | DM | 75 | 69 | (*S17*) |
| OD + X | $OD + O_3 \rightarrow DO_2 + O_2$ | $(7.3 \pm 1.1) \times 10^{-14}$ | DMH | 1 | 1 | (*S1*),(*S18*) |
| | $OD + D_2 \rightarrow D_2O + D$ | $(1.65 \pm 0.13) \times 10^{-15}$ | DM | 1 | 1 | (*S19*) |
| | $OD + OD \rightarrow D_2O + O$ | $(4.34 \pm 0.63) \times 10^{-13}$ | DM | 0 | 0 | (*S20*) |
| | $OD + DO_2 \rightarrow D_2O +$ | $(3.8 \pm 0.9) \times 10^{-11}$ | DM | 0 | 0 | (*S21*) |
| | $O_2$ | | | | | |
| | $OD + D_2O_2 \rightarrow D_2O + DO_2$ | $(5.91 \pm 0.42) \times 10^{-13}$ | DM | 0 | 0 | (*S22*) |
| | $OD + CO \rightarrow D + CO_2$ | $(5.6 \pm 0.2) \times 10^{-14}$ | DM | 28 | 31 | (*S23-26*) |
| | **$OD + CO + N_2 \rightarrow DOCO + N_2$** | **$(1.1 \pm 0.4) \times 10^{-32}$** | | | | **(this work)** |
| | **$OD + CO + CO \rightarrow DOCO + CO$** | **$(1.7 \pm 0.7) \times 10^{-32}$** | | | | **(this work)** |
| | $OD + OD \rightarrow D_2O_2$ | Termolecular: $k_0^{300} = 6.9 \times 10^{-31}$ $k_\infty^{300} = 2.6 \times 10^{-11}$ | DMH | 0 | 0 | (*S1*) |
| | $O + OD \rightarrow O_2 + D$ | $(3.3 \pm 0.5) \times 10^{-11}$ | DMH | 7 | 8 | (*S1*) |
| D + X | $D + O_3 \rightarrow OD + O_2$ | $(2.9 \pm 0.3) \times 10^{-11}$ | DMH | 33 | 45 | (*S1*) |
| | $D + DO_2 \rightarrow OD + OD$ | $(7.2 \pm 1.4) \times 10^{-11}$ | DMH | 0 | 0 | (*S1*) |
| | $D + DO_2 \rightarrow O + D_2O$ | $(1.6 \pm 0.8) \times 10^{-12}$ | DMH | 0 | 0 | (*S1*) |
| | $D + DO_2 \rightarrow D_2 + O_2$ | $(6.9 \pm 2.8) \times 10^{-12}$ | DMH | 0 | 0 | (*S1*) |
| | $D + O_2 \rightarrow DO_2$ | Termolecular: $k_0^{300} = 4.4 \times 10^{-32}$ $k_\infty^{300} = 7.5 \times 10^{-11}$ | DMH | 0 | 0 | (*S1*) |
| DOCO + X | $DOCO + D \rightarrow D_2O + CO$ | $1.39 \times 10^{-11}$ | TH | 0 | 2 | (*S27*) |
| | $DOCO + O_3 \rightarrow OD + CO_2 + O_2$ | $(4 \pm 0.4) \times 10^{-11}$ (overall fit) | FIT | 2 | 45 | **(this work)** |
| | $DOCO + D \rightarrow D_2 + CO_2$ | $9.31 \times 10^{-11}$ | TH | 1 | 14 | (*S27*) |
| | $DOCO + OD \rightarrow D_2O + CO_2$ | $1.03 \times 10^{-11}$ | TH | 0 | 3 | (*S29*) |
| | $DOCO + O \rightarrow OD + CO_2$ | $1.44 \times 10^{-11}$ | TH | 0 | 8 | (*S30*) |
| | $DOCO + O_2 \rightarrow CO_2 + DO_2$ | $(1.9 \pm 0.2) \times 10^{-12}$ | DMH | 0 | 3 | (*S31*) |
| | DOCO LOSS | (fitted for each trace) | | | | |
| O + X | $O + O_3 \rightarrow O_2 + O_2$ | $(8.0 \pm 0.8) \times 10^{-15}$ | DM | 0 | 0 | (*S1*) |
| | $O + DO_2 \rightarrow OD + O_2$ | $(5.9 \pm 0.3) \times 10^{-11}$ | DMH | 1 | 0 | (*S1*) |
| | $O + D_2O_2 \rightarrow OD + DO_2$ | $(1.7 \pm 0.3) \times 10^{-15}$ | DMH | 0 | 0 | (*S1*) |
| $DO_2$ + X | $DO_2 + O_3 \rightarrow OD + O_2 + O_2$ | $(1.9 \pm 0.3) \times 10^{-15}$ | DMH | 0 | 0 | (*S1*) |

[1] The source of the value used in the model is indicated: (TH) indicates a theoretical value, (DMH) indicates a direct experimental measurement of the Hydrogen-substituted reaction, and (DM) indicates a direct experimental measurement. (FIT) indicates a globally fitted reaction rate, specifically $DOCO + O_3$, which provided the best fit at $(4.0\pm0.4)\times10^{-11}$ $cm^3$ $molecule^{-1}$ $s^{-1}$ for all scans.

***Notes on Specific Reaction Rates:***

**<u>$OD + CO \rightarrow D + CO_2$</u>** This reaction rate was determined from the weighted average of three measurements from Paraskevopoulos *et al.* (*S24*), Golden *et al.* (*S26*), and Westenberg *et al.* (*S25*). These values are $k_1 = (5.2\pm0.5)\times10^{-14}$, $(6.6\pm0.4)\times10^{-14}$, and $(5.48\pm0.2)\times10^{-14}$ $cm^3$ $molecules^{-1}$ $s^{-1}$, respectively. The result of the weighted average is $k_1 = (5.6\pm0.2)\times10^{-14}$ $cm^3$ $molecules^{-1}$ $s^{-1}$.

***Sensitivity Analysis:*** In order to determine the sensitivity of the model on each of these parameters, a sensitivity analysis was done, measuring the variation of the DOCO and OD peak concentrations with each of the rate constants. The sensitivity, $S_{DOCO}$, is defined for a given rate constant $k$ as

$$S_{DOCO} = \frac{k}{[\mathrm{DOCO}]_{\max}} \frac{\partial [\mathrm{DOCO}]_{\max}}{\partial k},$$

where $[\mathrm{DOCO}]_{\max}$ is the maximum concentration of DOCO. $S_{OD}$ is defined in a similar manner. These values essentially represent the fractional fluctuation of [DOCO] or [OD] with a fractional change in $k$. Values of these parameters are given in Table S2 for the conditions

$[O_3] = 1\times10^{15}$ molecules cm$^{-3}$

$[D_2] = 1\times10^{17}$ molecules cm$^{-3}$

$[CO] = 1\times10^{17}$ molecules cm$^{-3}$

$[N_2] = 1\times10^{18}$ molecules cm$^{-3}$.

***Results of rate equation model fits:*** To compare the rate equation model to the experimental data, two parameters were fitted for each OD(t), DOCO(t) trace: an overall scaling factor for both OD and DOCO and a DOCO loss rate. Additionally, we found it necessary to fix the DOCO+$O_3$ rate as a constant and shared parameter for all traces, which yields a rate constant of $4\times10^{-11}$ cm$^3$ molecule$^{-1}$ s$^{-1}$. The scaling factor accounts for uncertainties in both the effective optical path length and the OD* chemistry involved that establishes the initial OD concentration measured at steady-state. The results of these fits as a function of CO are shown in Fig. S10A-B. The averaged values for the overall scaling factor and the DOCO loss rate are 0.14±0.05 and $(4.7\pm0.7)\times10^3$ s$^{-1}$, respectively. We observe a 65% correlation between the fitted overall scaling factor and DOCO loss rate.

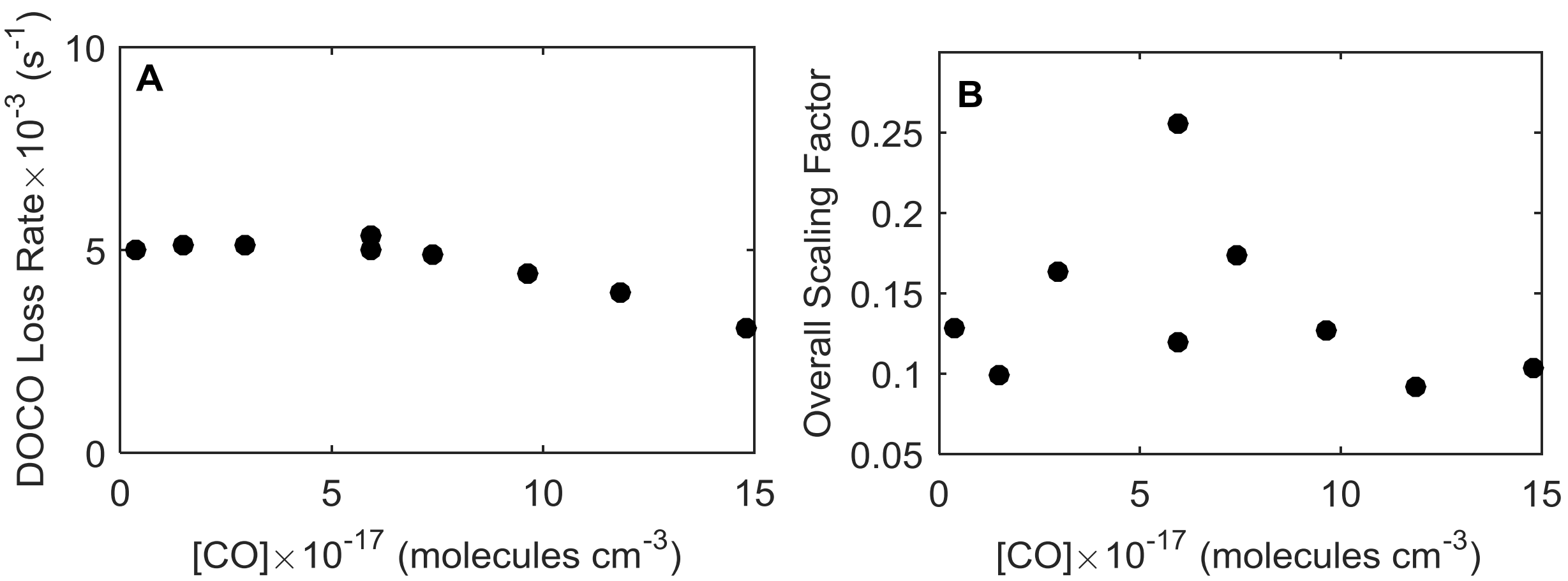

**Fig. S10**. Results of fitting the DOCO loss rate **(A)** and overall scaling factor **(B)** in the model to the data.

§5. Table of Statistical and Systematic Errors

**Table S3:** Summary of statistical and systematic errors

| | | **Error Source** | $\boldsymbol{k_{1a}^{(CO)}}$ | $\boldsymbol{k_{1a}^{(N2)}}$ |
|---|---|---|---|---|
| **Statistical Errors** | | (statistical, from fit residual) | 6% | 10% |
| **Experimental Control** | §1 | Flow & Pressure Measurement | 7% (stat) | |
| **Molecular Parameters** | §2 | OD Cross Section | 10% (stat) | |
| | §2 | DOCO Cross Section | 10% (stat) | |
| **Secondary Reactions** | §3 | Effect of $D_2$ | 8% (stat) | |
| | §3 | Effect of $O_3$ | 11% (stat) | |
| **Data Analysis** | §3 | Cross-contamination of OD and $D_2O$ | -1% (sys) | |
| | §3 | Effect of Integration Time | +9% (sys), 21% (stat) | |
| | §3 | OD Vibrational Excitation | -10% (sys) | |
| | | | | |
| | | Total Systematic Error Budget | (-11%,+9%) | |
| | | Total Statistical Error Budget | 28% | 29% |
| | | Total Error Budget | (-39%,+37%) | (-40%,+38%) |